\documentclass[journal]{IEEEtran}
\usepackage{amssymb,amsmath,amsthm, mathtools}
\usepackage{graphicx}
\usepackage{cite}
\usepackage{latexsym}
\usepackage{verbatim}
\usepackage{amsmath}
\usepackage{amssymb}
\usepackage{caption}
\usepackage{url}
\usepackage{epstopdf}
\usepackage{algorithm}
\usepackage{algcompatible}
\usepackage[noend]{algpseudocode}
\usepackage{times}
\usepackage{stfloats}
\usepackage{multirow}
\usepackage{enumerate}
\usepackage{xspace}
\usepackage{longtable}
\usepackage{mathrsfs}
\usepackage{mathtools}
\usepackage{algcompatible}
\usepackage{fancyhdr}
\usepackage[noend]{algpseudocode}
\usepackage{amssymb}% http://ctan.org/pkg/amssymb
\usepackage{pifont}% http://ctan.org/pkg/pifont
\usepackage{caption}
\usepackage{setspace}

\usepackage{color}

\usepackage{subcaption}
\usepackage{float}
\newcommand{\TVBD}{\ensuremath {\mathit{TVBD}}{\xspace}}
\newcommand{\fc}{\ensuremath {\mathit{FC}}{\xspace}}
\newcommand{\fcc}{\ensuremath {\mathit{FCC}}{\xspace}}
\newcommand{\crn}{\ensuremath {\mathit{CRN}}{\xspace}}

\newcommand{\nch}{\ensuremath {\mathit{s}}{\xspace}} % number of tv channels to be shared
\newcommand{\su}{\ensuremath {\mathit{SU}}{\xspace}}
\newcommand{\pu}{\ensuremath {\mathit{PU}}{\xspace}}

\newcommand{\pir}{\ensuremath {\mathit{PIR}}{\xspace}}

\newcommand{\hmac}{\ensuremath {\mathit{HMAC}}{\xspace}}

\newcommand{\rsa}{\ensuremath {\mathit{N}}{\xspace}}

\newcommand{\size}{\ensuremath {\mathit{\sigma}}{\xspace}}

\newcommand{\perc}{\ensuremath {\mathit{\varrho}}{\xspace}}
\newcommand{\chn}{\ensuremath {\mathit{chn}}{\xspace}}
\newcommand{\ckf}{\ensuremath {\mathit{CF}}{\xspace}} % Cuckoo Filter
\newcommand{\chr}{\ensuremath {\mathit{char}}{\xspace}} % Characteristics of the SU's device
\newcommand{\param}{\ensuremath {\mathit{par}}{\xspace}} % Values pf the different parameters of the SU's device
\newcommand{\db}{\ensuremath {\mathit{DB}}{\xspace}} % Spectrum database
\newcommand{\qr}{\ensuremath {\mathit{query}}{\xspace}} % Spectrum database
\newcommand{\resp}{\ensuremath {\mathit{resp}}{\xspace}} %  Database response to the query
\newcommand{\qp}{\ensuremath {\mathit{QS}}{\xspace}} % Spectrum database

\newcommand{\fp}{\ensuremath {\mathit{\epsilon}}{\xspace}}
\newcommand{\ts}{\ensuremath {\mathit{ts}}{\xspace}} %  Database response to the query
\newcommand{\rw}{\ensuremath {\mathit{r}}{\xspace}} %  Size of response
\newcommand{\cl}{\ensuremath {\mathit{c}}{\xspace}} %  number of columns in the response
\newcommand{\row}{\ensuremath {\mathit{row}}{\xspace}} %  One row of the response
\newcommand{\x}{\ensuremath {\mathit{x}}{\xspace}} %  String representing one row of the response
\newcommand{\y}{\ensuremath {\mathit{y}}{\xspace}} %  String constructed by the user for lookup
\newcommand{\avl}{\ensuremath {\mathit{avl}}{\xspace}} %  x location coordinate
\newcommand{\lx}{\ensuremath {\mathit{locX}}{\xspace}} %  x location coordinate
\newcommand{\ly}{\ensuremath {\mathit{locY}}{\xspace}} %  y location coordinate
\newcommand{\ncells}{\ensuremath {\mathit{m}}{\xspace}} %  number of cells in PriSpectrum protocol

\newcommand{\LPDB}{\ensuremath {\mathit{LPDB}}{\xspace}}
\newcommand{\LPDBQS}{\ensuremath {\mathit{LPDBQS}}{\xspace}}

\newcommand{\PrSpec}{\ensuremath {\mathit{PriSpectrum}}{\xspace}}

\newcommand{\simu}{\ensuremath{\mathcal{H}}}
\newcommand{\algorithmicbreak}{\textbf{break}}
\newcommand{\BREAK}{\STATE \algorithmicbreak}
\makeatother
{\bfseries}{\rmfamily}
\newtheorem{assumption}{Security Assumption}{\bfseries}{\rmfamily}
{\bfseries}{\rmfamily}
\newtheorem{mytheorem}{Theorem}{\bfseries}{\rmfamily}
\renewcommand{\figurename}{Figure}

\begin{document}

\title{Location Privacy Preservation in Database-driven Wireless Cognitive Networks through Encrypted Probabilistic Data Structures}
\vspace{-0.4in}
\author{Mohamed Grissa, Attila A. Yavuz, and Bechir Hamdaoui\\
\small Oregon State University, grissam,attila.yavuz,hamdaoui@oregonstate.edu\\
\vspace{-15pt}
\thanks{This work was supported in part by the US National Science Foundation under NSF award CNS-1162296. This manuscript is an extension of~\cite{grissa2015cuckoo}, published in: Computer Networks and Information Security (WSCNIS), 2015 World Symposium on. The authors are with Oregon State University, Corvallis, OR 97331
USA (e-mail: grissam@oregonstate.edu; attila.yavuz@oregonstate.edu;
hamdaoui@oregonstate.edu).}
\thanks{\copyright~2017 IEEE. Personal use of this material is permitted. Permission from IEEE must be obtained for all other uses, in any current or future media, including reprinting/republishing this material for advertising or promotional purposes, creating new collective works, for resale or redistribution to servers or lists, or reuse of any copyrighted component of this work in other works.}
}
%\IEEEoverridecommandlockouts
%\IEEEpubid{\makebox[\columnwidth]{978-1-4799-9907-1/15/\$31.00~
%\copyright2015
%IEEE \hfill} \hspace{\columnsep}\makebox[\columnwidth]{ }}
\maketitle
{\let\thefootnote\relax\footnote{{\\Digital Object Identifier 10.1109/TCCN.2017.2702163}}}
%%%%%%%%%%%%%%%%%%%%%%%%%%%%%%%%%%%%%%%%%%%%%%%%%%%%%%%%%%%%%%%%%%%%%
%%%%%%%%%%%%%%%%%%%%%%%%%%%%%% Abstract %%%%%%%%%%%%%%%%%%%%%%%%%%%%%
%%%%%%%%%%%%%%%%%%%%%%%%%%%%%%%%%%%%%%%%%%%%%%%%%%%%%%%%%%%%%%%%%%%%%
\begin{abstract}
In this paper, we propose new location privacy preserving
schemes for database-driven cognitive radio networks (\crn s) that protect secondary users' (\su s) location privacy while allowing them to learn spectrum availability in their vicinity. Our schemes harness probabilistic set membership data structures to exploit the structured nature of spectrum databases (\db s)
and \su s' queries. This enables us to create a compact representation
of \db~that could be queried by \su s without having to share their location with \db, thus guaranteeing their location privacy. Our proposed schemes offer different cost-performance characteristics. Our first scheme relies on a simple yet powerful two-party protocol that achieves unconditional security with a plausible communication overhead by making \db~send
a compacted version of its content to \su~which needs only to
query this data structure to learn spectrum availability. Our
second scheme achieves significantly lower communication and
computation overhead for \su s, but requires an additional architectural entity which receives the compacted version of the database and fetches the spectrum
availability information in lieu of \su s to alleviate the overhead
on the latter. We show that our schemes are secure, and also
demonstrate that they offer significant advantages over existing
alternatives for various performance and/or security metrics.
\end{abstract}

\begin{IEEEkeywords}
Database-driven spectrum availability, location privacy preservation, cognitive radio networks, set membership data structures.
%\vspace{-10mm}
\end{IEEEkeywords}

%%%%%%%%%%%%%%%%%%%%%%%%%%%%%%%%%%%%%%%%%%%%%%%%%%%%%%%%%%%%%%%%%%%%%
%%%%%%%%%%%%%%%%%%%%%%%%%%%% Introduction %%%%%%%%%%%%%%%%%%%%%%%%%%%
%%%%%%%%%%%%%%%%%%%%%%%%%%%%%%%%%%%%%%%%%%%%%%%%%%%%%%%%%%%%%%%%%%%%%
\section{Introduction}
\label{sec:Introduction}
Cognitive radio networks ($\crn$s) have emerged as a key technology for addressing the problem of spectrum utilization inefficiency~\cite{FCC2002Spectrum, khalfi2014optimal,zhu2016you,guizani2015large,adem2015impact,khalfi2015distributed,adem2014delay}. $\crn$s allow unlicensed users, also referred to as {\em secondary users (\su s)}, to access licensed frequency bands opportunistically, so long as doing so does not harm licensed users, also referred to as {\em primary users (\pu s)}.
In order to enable $\su$s to identify vacant frequency bands, also called white spaces, the federal communications commission (\fcc) has adopted two main approaches: {\em spectrum sensing-based approach} and {\em geo-location database-driven approach}.

In the sensing-based approach~\cite{wang2014location}, \su s themselves sense the licensed channels to  decide whether a channel is available prior to using it so as to avoid harming \pu s. In the database-driven approach, \su s rely on a geo-location database (\db) to obtain channel availability information. For this,
\su s are required to be equipped with GPS devices so as to be able to query \db~on a regular basis using their exact locations. Upon receipt of a query, \db~returns to \su~the list of available channels in its vicinity, as well as the transmission parameters that are to be used by \su. This database-driven approach has advantages over the sensing-based approach. First, it pushes the responsibility and complexity of complying with spectrum policies to \db. Second, it eases the adoption of policy changes by limiting updates to just a handful number of databases, as opposed to updating large numbers of devices~\cite{zhu2015protocol}.

Companies, like Google and Microsoft, are selected by FCC to administrate these geo-location databases, following the guidelines provided by {\em PAWS (Protocol to Access White-Space)}~\cite{zhu2015protocol}. {\em PAWS} protocol defines guidelines and operational requirements for both the spectrum database and the \su s querying it. These requirements include: \su s need to be equipped with geo-location capabilities, \su s must query \db~with their specific location to check channel availability before starting their transmissions, \db~must register \su s and manage their access to the spectrum, \db~must respond to \su s' queries with the list of available channels in their vicinity along with the appropriate transmission parameters.
% Database-based \crn s are also proven to have several other advantages compared to other spectrum sharing approaches~\cite{gurney2008geo}.

%\footnote{
%{\em PAWS} is a protocol introduced to enable interoperability between devices and databases~\cite{zhu2015protocol}.}
Despite their effectiveness in improving spectrum utilization efficiency, database-driven $\crn$s suffer from serious security and privacy threats. The disclosure of location privacy of $\su$s has been one of such threats to \su s when it comes to obtaining spectrum availability from \db s. This is simply because \su s have to share their locations with \db~to learn about spectrum availability. The fine-grained location, when combined with publicly available information, can lead to even greater private information leakage. For example, it can be used to infer private information like shopping patterns, preferences, behavior and beliefs, etc.~\cite{wicker2012loss}. Being aware of such potential privacy threats, $\su$s may refuse to rely on \db~for spectrum availability information. Therefore, there is a critical need for location-privacy preserving schemes for database-driven spectrum access.

    \subsection{Our Contribution}
In this paper, we propose two location privacy-preserving
schemes for database-driven \crn s with different performance and architectural benefits. The first scheme, {\em location privacy in database-driven CRNs} (\LPDB), provides optimal location privacy to \su s within \db's coverage area by leveraging {\em set membership data structures} (used to test whether an element is a member of a set) to construct a compact version of \db. The second scheme, {\em \LPDB~with two servers} (\LPDBQS), minimizes the overhead at \su's side at the cost of deploying an additional entity in the network. The cost-performance tradeoff gives more options to system designers to decide which topology and which approach is more suitable to their specific requirements.

Both approaches exploit two important facts: (i) Spectrum databases are highly structured~\cite{zhu2015protocol}; and (ii) \su s queries contain always the same device-specific characteristics (e.g., device type, antenna hight, frequency range, etc.)~\cite{zhu2015protocol}.
The highly structured property of the database refers to the fact that \db's structure is usually agreed upon by the FCC and the database administrators, like Google, Microsoft, etc, and that the queries and messages exchanged by \db~and \su s have a specific format in terms of what data they include. This well-defined information is available to both database administrators and \su s which allows them to have an idea on what kind of data the other party will include in its query/response, and also to compact both \db's content and the queries using probabilistic data structures. In fact, and as recommended by the {\em PAWS} standard, the database should always reply to \su s with a set of predetermined information. This allows \db~to compact its content to include only this information, which significantly reduces queried data sizes, and enables \su s to emulate \db's response when querying the probabilistic data structure as we show next.

A desirable property of our schemes is their simplicity that is expected to facilitate their applicability in real-life applications. Our proposed schemes offer various cost-performance trade-offs that can meet the requirements of different applications. We study these tradeoffs and show that high privacy and better performance for \su s' can be achieved, but at the cost of deploying an additional architectural entity in the system. We show that our proposed schemes are secure and more efficient than their existing counterparts. In addition, we study the impact of system parameters on the performances of our proposed schemes, and compare them against those obtained via existing approaches.

Compared to our preliminary work~\cite{grissa2015cuckoo}, this paper provides: (i) A new scheme, \LPDBQS, with multiple deployment scenarios, that improves the overhead on \su s' side by relying on an extra architectural entity; (ii) An improvement to our previously proposed scheme, \LPDB, by incorporating spectrum sensing to reduce the impact of the false positive rate of the used {\em set membership data structure} on spectrum availability information's accuracy; (iii) A detailed security analysis of the proposed schemes; and (iv) More detailed performance analysis with more evaluation metrics.

The remainder of this paper is organized as follows: We discuss related work in Section~\ref{sec:related}. We present our system and threat models along with our security assumptions in Section~\ref{sec:Preliminaries}. Section~\ref{sec:cuckoo} provides a brief overview of the {\em set membership data structure} that we use in this paper. In Section~\ref{sec:ProposedScheme}, we present our first scheme \LPDB. We describe our second scheme \LPDBQS~in Section~\ref{sec:secondScheme}. We evaluate and analyze the performance of the proposed schemes in Section~\ref{sec:PerformanceComp}, and conclude our work in Section~\ref{sec:Conclusion}.

%%%%%%%%%%%%%%%%%%%%%%%%%%%%%%%%%%%%%%%%%%%%%%%%%%%%%%%%%%%%%%%%%%%%%
%%%%%%%%%%%%%%%%%%%%%%%%%%%% Related Work %%%%%%%%%%%%%%%%%%%%%%%%%%%
%%%%%%%%%%%%%%%%%%%%%%%%%%%%%%%%%%%%%%%%%%%%%%%%%%%%%%%%%%%%%%%%%%%%%
\section{Related Work}
\label{sec:related}
Despite its importance, the location privacy issue in \crn s only recently gained interest from the research community~\cite{grissa2017location}. Some works focused on addressing this issue in the context of collaborative spectrum sensing~\cite{li2012location,grissa2015location,wangprivacy,grissa2016efficient,grissa2017preserving} while others focused on addressing it in the context of dynamic spectrum auction~\cite{liu2013location}. 
However, these works are not within the scope of this paper as we focus on the location privacy issue in database-driven \crn s.

Protecting \su s' location privacy in database-driven \crn s is a very challenging task, since \su s are required to provide their physical locations to \db~in order for them to be able to learn about spectrum opportunities in their vicinities. Recently developed techniques mostly adopt either the {\em k-anonymity}~\cite{gruteser2003anonymous}, {\em Private Information Retrieval (\pir)}~\cite{chor1998private}, or {\em differential privacy}~\cite{dwork2008differential} concepts. However, direct adaptation of such concepts yield either insecure or extremely costly results.
For instance, {\em k-anonymity} guarantees that \su's location is indistinguishable among a set of $k$ points, which could be achieved through the use of dummy locations by generating $k-1$ properly selected dummy points, and performing $k$ queries to \db~using both the real and dummy
locations. For example, Zhang et al.~\cite{zhang2015optimal} rely on this concept to make each \su~query \db~by sending a square cloak region that includes its actual location. Their approach makes a tradeoff between providing high location privacy and maximizing some utility, which makes it suffer from the fact that achieving a high
location privacy level results in a decrease in spectrum utility.

\pir, on the other hand, allows a client to obtain information from a database while preventing the database from learning which data is being retrieved. Several approaches have used this approach. For instance, Gao et al.~\cite{gao2013location} propose a \pir-based approach, termed \PrSpec, that relies on the \pir~scheme of Trostle et al.~\cite{trostle2010efficient} to defend against a newly identified attack that exploits spectrum utilization pattern to localize \su s. Troja et al.~\cite{troja2014leveraging,troja2015efficient} propose two other \pir-based approaches that try to minimize the number of \pir~queries by either allowing \su s to share their availability information with other \su s~\cite{troja2014leveraging} or by exploiting trajectory information to make \su s retrieve information for their current and future positions in the same query~\cite{troja2015efficient}.
Despite their merit in providing location privacy to \su s these \pir-based approaches incur high overhead especially in terms of computation.

Using {\em differential privacy}, Zhang et al.~\cite{zhang2015achieving} rely on the {\em $\epsilon$-geo-indistinguishability} mechanism~\cite{andres2013geo} to make \su s obfuscate their location. However, such a mechanism introduces noise to \su's location which may impact the accuracy of the spectrum availability information retrieved.

There have also been other privacy-enhancing technologies (PETs) that are not specific to \crn s but are designed to enable private queries over a database in general. However, many of these PETs are designed for applications that do not fit in the context of \crn s. For instance, oblivious random access memory (ORAM)~\cite{goldreich1996software} aims to enable a user to outsource its encrypted data to a database and to offer him/her the possibility to access this data while hiding the access patterns from the database~\cite{grissa2017location}. Searchable symmetric encryption~\cite{curtmola2011searchable} is another PET that is largely deployed to privately outsource one's data to another party
while maintaining the ability to selectively search over it~\cite{grissa2017location}. These PETs are designed for protecting queries and searches over data that is outsourced to a database, which is completely different from the \crn~scenario where the queried data belong to the database itself.

%%%%%%%%%%%%%%%%%%%%%%%%%%%%%%%%%%%%%%%%%%%%%%%%%%%%%%%%%%%%%%%%%%%%%
%%%%%%%%%%%%%%%%%%%%%%%%%%%% Preliminaries %%%%%%%%%%%%%%%%%%%%%%%%%%%
%%%%%%%%%%%%%%%%%%%%%%%%%%%%%%%%%%%%%%%%%%%%%%%%%%%%%%%%%%%%%%%%%%%%%
\section{System Model and Security Assumptions}
\label{sec:Preliminaries}
%In this section, we describe our system and security models.

\subsection{Database-driven $\crn$ Model}
We first consider a $\crn$ that consists of a set of \su s and a geo-location database (\db). \su s are assumed to be enabled with GPS and spectrum sensing capabilities, and to have access to \db~to obtain spectrum availability information within its operation area. To learn about spectrum availability, a \su~queries \db~by including its location and its device characteristics. \db~responds with a list of available channels at the specified location and a set of parameters for transmission over those channels. \su~then selects and uses one of the returned channels. While using the channel, \su~needs to recheck its availability on a daily basis or whenever it changes its location by $100$ meters as mandated by {\em PAWS}~\cite{zhu2015protocol}.

We then investigate incorporating a third entity to the network along with \db~and \su s. This entity, referred to as {\em query server} (\qp), has a dedicated high throughput link with \db. \qp~is used to guarantee computational location privacy while reducing the computational and communication overhead especially on \su s' side.

\subsection{Security Model and Assumptions}

\db~and~\qp~are assumed to be honest but curious. That is, \db~and \qp~follow the protocol honestly but may try to infer information on the input of other parties beyond what the output of the protocol reveals. Specifically, our objective is to prevent these two entities from learning \su s' location. Therefore, our security assumptions are as follows:

\begin{assumption} \label{secAssumption1} \db~and \qp~do not modify the integrity of their input. That is, (i) \db~does not maliciously change \su's query's content; (ii) \qp~does not modify the input that it receives from \db~or \su.
\end{assumption}

\begin{assumption}\label{secAssumption2}
\db~and \qp~do not collude with each other to infer the location of \su s from their queries.
\end{assumption}

We further assume that the communication between different entities is secured by a cryptographic protocol like TLS~\cite{dierks1999tls} as suggested by {\em PAWS}~\cite{zhu2015protocol}. This eliminates the risk of an adversary trying to eavesdrop the communication.

%%%%%%%%%%%%%%%%%%%%%%%%%%%%%%%%%%%%%%%%%%%%%%%%%%%%%%%%%%%%%%%%%%%%%
%%%%%%%%%%%%%%%%%%%%%%%%%%%% Preliminaries %%%%%%%%%%%%%%%%%%%%%%%%%%%
%%%%%%%%%%%%%%%%%%%%%%%%%%%%%%%%%%%%%%%%%%%%%%%%%%%%%%%%%%%%%%%%%%%%%
\section{Set Membership Data Structures}
\label{sec:cuckoo}
Our proposed privacy-preserving schemes utilize {\em set membership data structures} to exploit the highly structured property of \db. There are several data structures that are  designed for set membership tests, e.g. {\em bloom filter}~\cite{bloom1970space}, {\em cuckoo filter}~\cite{fan2014cuckoo}, etc. However, in this paper, we opt for {\em cuckoo filter} as the building block of our schemes. We use {\em cuckoo filter} to construct a compact representation of the spectrum geo-location database as explained in Sections~\ref{sec:ProposedScheme} \&~\ref{sec:secondScheme}. What motivates our choice is that {\em cuckoo filter} offers the highest space efficiency among its current well known alternatives, such as {\em bloom filters}. Besides, it has been proven to be more efficient than these alternatives especially for large sets. Finally, the {\em cuckoo filter} enjoys fast $Lookup$ and $Insert$ operations that are beneficial to our schemes.

%This makes it promising in  \db~that may contain entries corresponding to spectrum availability for a location resolution reaching $50\;meters$.

\begin{figure}%[h!]
\center
    % double column
    \includegraphics[width=0.36\textwidth]{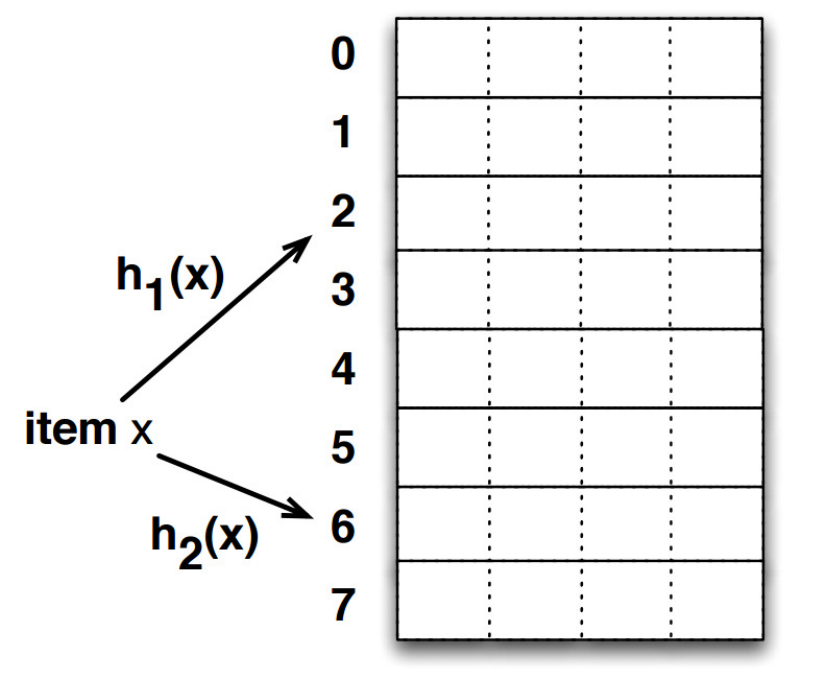}

    \caption{ Cuckoo Filter: 2 hashes per item, 8 buckets each containing 4 entries}
    \label{fig:ck}

\end{figure}

A cuckoo filter~\cite{fan2014cuckoo} uses {\em cuckoo hashing}~\cite{pagh2004cuckoo} and is designed to serve applications that need to store a large number of items while targeting low false positive rates and requiring storage space smaller than that required by bloom filters.
A false positive occurs when the membership test returns that an item exists in the {\em cuckoo filter} (i.e., belongs to the set) while it actually does not. A false negative, on the other hand, occurs when the membership test returns that an item does not exist while it actually exists.
In {\em cuckoo filters}, false positives are possible, but false negatives are not, and the target false positive rate, denoted throughout this paper by \fp, can be controlled but has a direct impact on the filter's size. Figure~\ref{fig:ck} shows an example of a {\em cuckoo filter} that uses two hashes per item and contains $8$ buckets each with $4$ entries.

A cuckoo filter has mainly two functions: An $Insert$ function that stores items in the filter, and a $Lookup$ function that checks whether an item exists in the filter. In the $Insert$ operation,  cuckoo filter stores a fingerprint $f$ of each item $x$, as opposed to storing the item itself. The space cost, in bits, of storing one item in the cuckoo filter using the $Insert$ function depends on the target false positive rate \fp~and is given by $(log_2(1/\fp)+ log_2(2\beta))/\alpha$ where $\alpha$~is the load factor of the filter which defines its maximum capacity, and $\beta$ is the number of entries/slots per bucket. Once the maximum feasible, $\alpha$, is reached, insertions are likely to fail, and hence, the filter must expand in order to store more items~\cite{fan2014cuckoo}. The $Lookup$ operation is performed by first computing a fingerprint of the desired item and two indexes, representing the potential locations (or buckets) of this item in the filter, and then checking whether these two locations contain the item.
%%%%%%%%%%%%%%%%%%%%%%%%%%%%%%%%%%%%%%%%%%%%%%%%%%%%%%%%%%%%%%%%%%%%%
%%%%%%%%%%%%%%%%%%%%%%%%%%%% Proposed Schemes %%%%%%%%%%%%%%%%%%%%%%%%%%%
%%%%%%%%%%%%%%%%%%%%%%%%%%%%%%%%%%%%%%%%%%%%%%%%%%%%%%%%%%%%%%%%%%%%%
\section{Proposed Schemes}
In this section, we describe our proposed schemes. The first scheme, \LPDB, is simple as it involves only two parties, \su s and \db, and provides unconditional location privacy to \su s within the coverage area of \db. The second scheme, \LPDBQS, offers computational privacy with a significantly reduced overhead on \su s' side compared to \LPDB, but at the cost of introducing an extra architectural entity.

Since we are unable to access the actual spectrum database, we relied on two sources to have an estimate of this structure: First, we have relied on the recommendation of the PAWS standard~\cite{zhu2015protocol}, which defines the interaction between \su s and \db~and what information they should exchange. Second, we used graphical web interfaces provided to the public by white space database operators like Google~\cite{google}, Microsoft~\cite{microsoft}, iconectiv~\cite{iconectiv}, etc. These web interfaces comply with PAWS recommendation and allow an interested user to specify a location of interest and learn spectrum availability in that location to emulate the interaction between a \su~and \db~in real world. While the purpose of these interfaces was initially to provide a working platform as a showcase for FCC to acquire approval for operating spectrum database, we believe it has enough information to enable us to estimate the structure of the database and \su s' queries.

As required by PAWS, \su s must be registered with \db~to be able to query it for spectrum availability. Registered \su~starts by sending an initialization query to \db~which replies by informing the \su~of specific parameterized-rule values. These parameters include time periods beyond which the \su~must update its available-spectrum data, and maximum location change before needing to query \db~again. Afterwards, \su~queries \db~with an available spectrum query which contains its geolocation, device identifier, capabilities (to limit \db's response to only compatible channels) and antenna characteristics (e.g. antenna height and type). \db~then replies with the set of available channels in the \su's location along with permissible power levels for each channel.

Based on these interactions between \su~and \db, which we learned from {\em PAWS} and the database web interfaces, we estimate the structure of \db~to be as illustrated in Table~\ref{dbStruct}. Each row corresponds to a different combination of location pairs (\lx,\ly) and channel \chn. One location may contain several available channels at the same time. Note that even if the real structure deviates from the one illustrated in Table~\ref{dbStruct} (e.g. more/different attributes, more tables, etc), our schemes can be adapted to the new structure of both the queries and the database by designing or using a different probabilistic data structure(s). Also, even in this case, the {\em PAWS} standard requires that \db~always replies to spectrum availability queries with a set of predetermined values that have to be in the database no matter what structure it has. In that case, \db~only needs to insert these values in the cuckoo filter and this could be done independently from the database structure. %For simplicity, we will be using the structure illustrated in Table~\ref{dbStruct} throughout the paper.

\begin{table}[h!]
\centering
\caption{ Simplified example of \db's structure}
\label{dbStruct}

\renewcommand{\arraystretch}{1.25}{
% double column
\resizebox{.43\textwidth}{!}{
% single column
%\resizebox{.6\textwidth}{!}{
\begin{tabular}{c|c|c|c|c|c|c|c|c|}
\cline{2-9}
                               & \lx     & \ly     & \ts & \chn& \avl& $\param^1$& $\cdots$ & $\param^n$    \\ \hline
\multicolumn{1}{|c|}{$\row_1$} & $\lx_1$ & $\ly_1$ & t   & $\chn_1$ & $0$& $\param^1_1$& $\cdots$ & $\param^n_1$ \\ \hline
\multicolumn{1}{|c|}{$\row_2$} & $\lx_1$ & $\ly_1$ & t   & $\chn_2$ & $1$& $\param^1_2$& $\cdots$ & $\param^n_2$ \\ \hline
\multicolumn{1}{|c|}{$\vdots$ }  & $\vdots$  & $\vdots$  & $\vdots$  & $\vdots$ & $\vdots$& $\vdots$ & $\vdots$ &  $\vdots$       \\ \hline
\multicolumn{1}{|c|}{$\row_i$} & $\lx_2$ & $\ly_2$ & t   & $\chn_1$ & $1$& $\param^1_i$& $\cdots$ & $\param^n_i$ \\ \hline
\multicolumn{1}{|c|}{$\vdots$ }  & $\vdots$  & $\vdots$  & $\vdots$ & $\vdots$& $\vdots$ & $\vdots$  & $\vdots$ &    $\vdots$       \\ \hline
\multicolumn{1}{|c|}{$\row_r$} & $\lx_r$ & $\ly_r$ & t   &$\chn_1$ & $0$& $\param^1_r$& $\cdots$ & $\param^n_r$       \\ \hline
\end{tabular}}}
\center{\scriptsize{$\avl = 1$ means \chn~is available and $\avl = 0$ means \chn~is not available.}}

\end{table}

\subsection{\LPDB}
\label{sec:ProposedScheme}
In this section, we describe our basic scheme, which is referred to as {\em location privacy in database-driven $\crn$s} (\LPDB). The novelty of \LPDB~lies in the use of set membership data structures to construct a compact (space efficient) representation of \db~that can be sent to querying \su s~to inform them about spectrum availability.

In our scheme, instead of sending its location, a \su~sends only its characteristics (e.g., its device type, its antenna type, etc.), as specified by PAWS~\cite{zhu2015protocol}, to \db, which then uses them to retrieve the corresponding entries in all possible locations. \db~then puts these entries in a {\em cuckoo filter} and sends it to \su. Upon receiving this filter, \su~constructs a query that includes its characteristic information, its location, and one of the possible channels with its associated parameters. \su~then looks up this query in the received {\em cuckoo filter} to see whether that channel is available in its current location.

Parameters that are inserted in the response of \db~may include the location, time stamps, the available channels, and the transmission power to be considered when using those channels. \su's characteristics and \db~parameters could be agreed upon beforehand between \db~and \su s to make sure that \su~queries the {\em cuckoo filter} with the right parameters.

The proposed \LPDB~scheme is illustrated in Algorithm~\ref{alg1}, and briefly described as follows: First, each \su~starts by constructing $\qr$ to be sent to \db~by including a set of characteristics, which are specific to the  querying device, along with a time stamp \ts. \db~then retrieves the entries that correspond to \qr~and constructs a {\em cuckoo filter} \ckf~(which could be done offline). Since \db~contains availability status for each channel in each location, the number of entries satisfying \qr~will still be huge and one way to further reduce it is to retrieve only the information about available channels and ignore the other ones. Afterwards, \db~concatenates the data in each row to construct $x_{j}$ as illustrated in Step~\ref{xij2}, inserts it to \ckf~and sends \ckf~to \su.

\su~constructs a string $y$ by concatenating its location coordinates with a combination of one channel and its possible transmission parameters and tries to find whether $y$ exists in \ckf~by using the $Lookup$ operation of \ckf. \su~keeps changing the channel and the associated parameters until it finds the string $y$ in \ckf~or until \su~tries all possible channels. Note that, depending on the false positive rate \fp~of \ckf, even if the $Lookup$ operation returns $True$ it does not necessarily mean that the specified channel is available. Setting \fp~to be very small makes the probability of having such a scenario very small, thus reduces the risk of using a busy channel, but this cannot be done without increasing the size of \ckf. To further reduce the risk of falling into this case, we have also included an additional sensing step to confirm the query's result and give more accurate information about the status of the channel of interest. If \su~finds $y$ in \ckf, then it needs to sense the specific channel found in $y$ to confirm its availability. \su~can conclude that this channel is free and thus can use it only if the sensing result coincides with \ckf's response.

If, after trying all possible combinations, \su~does not find $y$ in \ckf, this means that no channel is available in the specified location as {\em cuckoo filters} do not incur any false negatives.

\begin{algorithm}[h!]
\caption{\LPDB~Algorithm}\label{alg1}
\begin{algorithmic}[1]

\State \su~queries \db~with $\qr \gets f(\chr, \ts)$; \label{alg1:query}
\State \db~retrieves $\resp$ containing $\rw$ entries satisfying $\qr$;
\State \db~constructs $\ckf$;
\For{\texttt{$j = 1,\ldots,\rw$}}
\If {$\avl_j = 1$}
\State $\x_{j} \gets (\lx_{j}\|\ly_{j}\|\chn\|\ts\|\ldots)$; \label{xij2}
\State \db~inserts $\x_{j}$ into $\ckf$: $\ckf.Insert(\x_{j})$;
\EndIf
\EndFor
\State \db~sends $\ckf$ to \su;
\State \su~initializes $decision$ $ \gets $ Channel is busy
\For{\texttt{all possible combinations of $\boldsymbol{\param}$}}
\State \su~computes $\y \gets (\lx \|\ly \|\chn_i \| \ts \|\ldots\| \param^n)$;
\If { $\ckf.Lookup(\y)$}
\State \su~senses $\chn$;
\If  {$Sensing(\chn)\gets$ available}
\State $decision$ $ \gets $ \chn~is available; \BREAK;
\EndIf
\EndIf
\EndFor

\Return $decision$
\end{algorithmic}
\end{algorithm}

When the size of \db~is not large, then \LPDB~ works well (as will be shown Section~\ref{sec:PerformanceComp}) by providing unconditional privacy with reasonably small amounts of overhead. However, a scalability issue may arise when the location resolution is very small (resolution used in \db~could be as small as 50 meters) and/or the area covered by \db~is large (e.g. at the scale of a country). In this case, the number of locations, and thus the number of entries in \db, can be large, and then even after relying on the {\em cuckoo filter}, the size of the data to be transmitted may still be impractical (e.g. in the order of gigabytes). This depends on the desired resolution and \db's covered area. Next, we present a discussion about a possible way to deal with this scalability issue in the case of a very large \db.

\subsubsection*{Performance-privacy tradeoff}

\begin{figure}[h!]
\center
	% double column
    \includegraphics[width=0.5\textwidth]{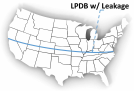}

    \caption{ Location Leakage}
    \label{fig:leakage}

\end{figure}

As discussed before, \LPDB~may suffer from a scalability issue when the size of \db's coverage area is very large. We can address this issue through the following observation. When the covered area is large and/or the location resolution is small, allowing \db~to learn one of \su's coordinates can drastically reduce the number of entries that \db~retrieves. This leads to considerably reduce the size of \ckf~to be transmitted, thus making the approach scalable. Interestingly, in the case of large areas, revealing one of \su's coordinates does not make it any easier for \db~to infer \su's location. To illustrate this, let's for example assume that \db~covers the entire surface of the United States, as shown in Figure~\ref{fig:leakage}. Allowing \db~to learn one coordinate (e.g the latitude) means that it can only learn that \su~is located somewhere on the blue line that spans the latitude of the whole country. But since \db~does not know the longitude of \su, then knowing the latitude only does not offer any practical information about \su's location.

 This, as shown in Section~\ref{performance}, drastically reduces the size of \ckf~transmitted by \db~at the cost of loosing the unconditional location privacy of \su s. However, when the coverage area of \db~is large, even revealing one of the coordinates still achieves high location privacy of \su s. Indeed, since databases (like those managed by Microsoft and Google) may cover an entire nation of the size of the United States, the revealed information is not sufficient to localize \su, yet, this reduces our scheme's overhead substantially.
The example of the United States in Figure~\ref{fig:leakage} shows that our scheme can offer high privacy even when one of the coordinates is revealed. Throughout, we refer to this variant of our scheme as \LPDB~with leakage.

It is worth reiterating that when the covered area is not too large, then the size of the {\em cuckoo filter} is practical and there is no need to reveal one of \su's coordinates. In this case, our scheme, \LPDB, provides unconditional privacy without incurring much overhead. The system regulator can decide about which approach to follow depending on the system constraints and the size \db's covered area.

%%%%%%%%%%%%%%%%%%%%%%%%%%%%%%%%%%%%%%%%%%%%%%%%%%%%%%%%%%%%%%%%%%%%%
%%%%%%%%%%%%%%%%%%%%%%%%%%%% Second Scheme %%%%%%%%%%%%%%%%%%%%%%%%%%%
%%%%%%%%%%%%%%%%%%%%%%%%%%%%%%%%%%%%%%%%%%%%%%%%%%%%%%%%%%%%%%%%%%%%%
\subsection{\LPDBQS}
\label{sec:secondScheme}

In this section, we propose a new scheme, \LPDBQS, which offers better
performance at \su s' side than that of \LPDB. This comes at the cost of deploying an additional entity, referred to as {\em query server} (\qp), and having a computational security as opposed to unconditional.  \qp~is introduced to handle \su s' queries instead of \db~itself, which prevents \db~from learning information related to \su s' location information. \qp~learns nothing but secure messages sent by \su s to check the availability of a specific channel.

$\bullet$ {\em Intuition}: We introduce \qp~to avoid sending \ckf, which might be large, to \su. Instead, \ckf, that contains \hmac~secure entries inserted by \db~using a secret key provided by \su, is sent to \qp~through a high throughput link pre-established with \db. \su~just needs to query, using \hmac~messages, \qp~which looks for its queries in \ckf. Using \hmac, \su~can hide the content of the query string, which includes its location information, among others, from \qp~which ignores the key used to construct the hashed query and the \ckf. This not only prevents \qp~from learning the query's content but also the entry that matches it in the filter. As most of the computation and communication overhead are incurred by both \db~and \qp, this scheme is the most efficient in terms of overhead incurred by \su s. \LPDBQS~is summarized in Algorithm~\ref{alg2} and described in the following.

\begin{algorithm}[h!]
\caption{\LPDBQS~Algorithm}\label{alg2}
\begin{algorithmic}[1]
\State \su~queries \db~with $\qr \gets f(k,\chr, \ts) $; \label{kchr}
\State \db~retrieves $\resp$ containing $\rw$ entries satisfying $\chr$;
\State \db~constructs $\ckf_k$;
\For{\texttt{$j = 1,\ldots,\rw$}}\label{for}
\If {$\avl_j = 1$}
\State $\x_{j} \gets (\lx_{j}\|\ly_{j}\|\ts\|\ldots\|\row_{j}(\cl))$;
%\State \db~inserts $\mathit{HMAC}_k(\x_{j})$ into $\ckf$;
\State $\ckf_k.Insert_{\scriptstyle \mathit{HMAC}_k}(\x_{j})$; \label{insert}
\EndIf
\EndFor
\State \db~sends $\ckf_k$ to \qp~over a high throughput link;\label{ckfk}
\State \su~initializes $decision$ $ \gets $ Channel is busy
\For{\texttt{all possible combinations of $\boldsymbol{\param}$}}
\State \su~computes $\y \gets (\lx \|\ly \|\chn \| \ts \|\ldots\| \param^n)$;
\State \su~computes $\y_k \gets \mathit{HMAC}_k(y)$ and sends it to \qp;\label{yk}
\State \qp~looks up for $\y_k$ in $\ckf_k$ using $Lookup$;
\If { $\ckf_k.Lookup(\y_k)$}
\State \su~senses $\chn$;
\If  {$Sensing(\chn)\gets$ available}
\State $decision$ $ \gets $ \chn~is available; \BREAK;
\EndIf
\EndIf
\EndFor

\Return $decision$
\end{algorithmic}
\end{algorithm}

 First, \su~starts by sending a secret key $k$, pre-established beforehand, along with its device characteristics to \db. \db~then retrieves only the entries that have available channels and that also comply with the device characteristics of the querying \su. Afterwards, \db~constructs a {\em cuckoo filter} $\ckf_k$ and inserts into it the entries retrieved in the previous step as shown in Steps~\ref{for}-\ref{insert}.
$Insert_{\scriptstyle \mathit{HMAC}_k}$, in Step~\ref{insert}, is a modified version of the $Insert$ procedure, where the fingerprint is replaced by an $\hmac_k$ function. \db~uses $\hmac_k$ with the secret key $k$, provided by \su, to construct hashed entries and insert them to $\ckf_k$. \db~then sends $\ckf_k$ to \qp~via the high speed link that connects \db~to~\qp.

\su~constructs a string $y$ by concatenating its location coordinates with a combination of one channel and its possible transmission parameters. Subsequently, \su~hashes $y$ using an \hmac~with the secret key $k$ and sends the new value $y_k$ to \qp~to find out whether $\ckf_k$ of \qp~contains $y_k$. If the query's combination is found in $\ckf_k$, then \su~needs to take one further step: It senses the channel that was included in the query. If the result of the sensing complies with the outcome of the $Lookup$ operation in $\ckf_k$, then \su~can conclude that this channel is available and, thus, it can use it for its future transmissions. In this case, \su~can stop querying \qp. The sensing operation is added to confirm the outcome of querying the {\em cuckoo filter} and overcome the risk of falling into the case of a false positive result that would eventually make \su~interfere with primary transmissions. In case the sensing result is different from the outcome of the $Lookup$ operation, then \su~keeps changing the channel and the associated parameters until \qp~finds $y_k$ in $\ckf_k$ or until \su~tries all possible channels and combinations.

This scheme considerably reduces the overhead perceived by \su s, as much of the computation is performed offline by \db, and  \su s do not need to download the {\em cuckoo filters} which are only sent to \qp~over a high throughput link.

If \db~knows the possible device characteristics of the querying \su s, this can help to further reduce the incurred overhead. Indeed, \db~can pre-compute several {\em cuckoo filters} for each possible combination of potential device parameters offline by relying on a set of secret keys $\mathcal{K}=\{k_1,\ldots,k_z\}$ that it generated beforehand. For each combination of parameters, \db~constructs multiple $\ckf_k$ with different keys from $\mathcal{K}$ to make sure that each \su~uses a different filter. \su s are not required to generate their own keys as in the previous variant. Whenever a \su~queries \db~for spectrum opportunities, \db~shares a secret key $k$ with it and sends the corresponding $\ckf_k$ to \qp. \su~uses $k$ to construct its hashed strings and query $\ckf_k$ of \qp~just like in Algorithm~\ref{alg2}.

%\noindent \textbf{Initialization:} This phase is executed only once at the beginning of the protocol or whenever the content of the database is updated. \db~starts by generating a set of secret keys $\mathcal{K}=\{k_1,\ldots,k_z\}$ then it constructs the {\em cuckoo filter} $\ckf$. Finally, \db~constructs $\ckf_{k_i}$s using \hmac~and the keys in $\mathcal{K}$. Now this allows \db~to have several copies of \ckf~ready for use whenever some \su s ask for spectrum availability information.
%
%\noindent \textbf{Private Query:} This is executed every time a \su~queries \db~for spectrum availability. As shown in Algorithm~\ref{alg3}, after receiving a query from a \su, \db~shares with it a secret key $k_l$ randomly picked from $\mathcal{K}$ and sends the corresponding {\em cuckoo filter} $\ckf_{k_l}$ to \qp. To learn about spectrum opportunities the \su~needs to query $\ckf_{k_l}$ at \qp~by following the procedure described in Lines~\ref{start}-\ref{end} of Algorithm~\ref{alg3}.

\subsubsection*{Leveraging a Secure Hardware}
As long as \db~and \qp~do not collude, as stated in Security Assumption~\ref{secAssumption2}, neither of them can infer the coordinates of \su s from the keyed one-way function output. To mitigate the non-collusion requirement between \fc~and \qp, \LPDBQS~could be implemented in a slightly different way by relying on a secure hardware (e.g., a secure co-processor or a trusted platform module) that can perform cryptographic operations without exposing its embedded private key. This hardware can be deployed inside \db~itself and play the role of \qp. Such a high-end secure hardware is physically shielded from penetration~\cite{yee1995secure}, and any tampering from the
adversary, potentially \db, triggers the automatic erasure of sensitive memory areas containing critical secrets~\cite{pub1999security}. When a secure hardware meets the FIPS 140-2 level 4~\cite{pub2001140} physical security requirements, it becomes infeasible for \fc~to tamper with the operations executed by this hardware. \db~sends the {\em cuckoo filters} to this hardware, and \su s have to query this hardware to learn about spectrum availabilities.
%

%%%%%%%%%%%%%%%%%%%%%%%%%%%%%%%%%%%%%%%%%%%%%%%%%%%%%%%%%%%%%%%%%%%%%
%%%%%%%%%%%%%%%%%%%%%%%%%%%% Security Analysis %%%%%%%%%%%%%%%%%%%%%%%%%%%
%%%%%%%%%%%%%%%%%%%%%%%%%%%%%%%%%%%%%%%%%%%%%%%%%%%%%%%%%%%%%%%%%%%%%
\section{Security Analysis}
\label{sec:secAnalysis}

In this section, we analyze the security of our proposed schemes \LPDB~and \LPDBQS.

%Let $(\LU,\LD,\LQ)$ be history lists that include the information that is learned by \su s, \db~and \qp, respectively, during the execution of our proposed protocols.

\begin{mytheorem}
Under Security Assumptions~\ref{secAssumption1} and \ref{secAssumption2}, \LPDB~does not leak any information on \su s' location.
\end{mytheorem}
\noindent {\em Proof:} We construct a history list \simu~of each entity's knowledge about \su s' information during the execution of \LPDB.

\noindent  ${\boldsymbol \su.}$ A \su~cannot learn anything about other \su s information nor the filters $\{\ckf_{i,t}\}_{i=1,t=t_0}^{n-1, t_f}$ that they receive from \db~as the communication between each \su~and \db~is secured, i.e. $\simu_\su = \emptyset$. Note that, even if, a \su~would learn the filters of other \su s, i.e. $\simu_\su = \{\ckf_{i,t}\}_{i=1,t=t_0}^{n-1, t_f}$, $\simu_\su$ includes no information about \su s' location.

\noindent ${\boldsymbol \db.}$ In Step~\ref{alg1:query} of Algorithm~\ref{alg1}, \db~learns $\simu_\db = \{\chr\}_{i=1}^{n}$ which contains the characteristics of the querying \su s. $\simu_\db$ may include information like frequency ranges in which \su~can operate, antenna characteristics, etc. This information is not related to the querying \su s' location. This shows that the knowledge that \db~gains during the execution of \LPDB~does not allow it to infer \su s' location when they try to learn about spectrum opportunities. \LPDB~offers an unconditional privacy in the sense that \db's knowledge about \su s' location, during the execution of \LPDB, does not increase compared to its initial knowledge, which is necessarily the coverage area of \db. \hfill$\square$

\begin{mytheorem}
Under Security Assumptions~\ref{secAssumption1} and \ref{secAssumption2} \LPDBQS~does not leak any information about \su s' location beyond $\kappa-\hmac$ secure values.
\end{mytheorem}

\noindent {\em Proof:} We construct a history list of each entity's knowledge during the execution of \LPDBQS.

\noindent  ${\boldsymbol \su.}$ As the communication between different entities is secured, \su s cannot learn any information about the communicated information of other entities, i.e. $\simu_\su = \emptyset$.

\noindent ${\boldsymbol \db.}$ In Line~\ref{kchr} of Algorithm~\ref{alg2}, \db~learns $\simu_\db = \{k_{i,t},\chr_i,\ts_t\}_{i=0,t=t_0}^{n,t_f}$. Obviously, \su s' secret keys $\{k_{i,t}\}_{i=0,t=t_0}^{n,t_f}$ and timestamp values $\{\ts_t\}_{t=t_0}^{t_f}$ cannot leak any information about \su s' location since these values are not correlated to their physical location. Similarly, their characteristics $\{\chr_i\}_{i=1}^n$ contain information about \su s' devices capabilities, like their possible transmit powers, antennas height, etc, which cannot be used to localize them. This proves that \db's~knowledge about \su s' location during the execution of \LPDBQS~does not differ from its initial knowledge; i.e. that \su s are within \db's covered area.

\noindent ${\boldsymbol \qp.}$ As indicated in Lines~\ref{ckfk} \& \ref{yk}~of Algorithm~\ref{alg2}, the only information that \qp~can learn during the execution of \LPDBQS, is $\simu_\qp = \{y_{k_{i,t}},\ckf_{k_{i,t}}\}_{i=1,t=t_0}^{n,t_f}$. $\{y_{k_{i,t}}\}_{i=1,t=t_0}^{n,t_f}$ are as secure as \hmac. The elements of $\{\ckf_{k_{i,t}}\}_{i=1,t=t_0}^{n,t_f}$  are computed using a pseudo random function (as an \hmac~is also a pseudo random function) with \su s' secret keys $\{k_{i,t}\}_{i=1,t=t_0}^{n,t_f}$, where $\{k_{i,t}\}_{i=1,t=t_0}^{n,t_f} \overset{\$}{\gets} \{0,1\}^\kappa$ and $\kappa$ is the security level. $\{y_{k_{i,t}}\}_{i=1,t=t_0}^{n,t_f}$ are independent from each other. The same applies to $\{\ckf_{k_{i,t}}\}_{i=1,t=t_0}^{n,t_f}$. Each query from $\{y_{k_{i,t}}\}_{i=1,t=t_0}^{n,t_f}$ has a corresponding \hmac~key, which means that even for the same \su~querying the same information, there will be randomly independent and uniformly distributed outputs generated by \db~and \su s.
Since only \su s~and \db~know the keys $\{k_{i,t}\}_{i=1,t=t_0}^{n,t_f}$ and that these keys are updated for every query made by \su s, \qp~cannot learn any information about \su s' location as long as it does not collude with \db~as stated in Security Assumption~\ref{secAssumption2}. Correlating queries $\{y_{k_{i,t}}\}_{i=1,t=t_0}^{n,t_f}$ to \su s' physical location is equivalent to breaking the underlying \hmac~or $PRF$, which is of probability $1/2^\kappa$.

 We can conclude that \LPDBQS~is as secure as the underlying \hmac.\hfill$\square$
%Note that the risk of an external attacker or an intruder that tries to eavesdrop the communication between a \su~and \db~in order to infer \su's location is eliminated by relying on a cryptographic protocol like TLS~\cite{dierks1999tls} as suggested by PAWS~\cite{zhu2015protocol,chen2015protocol}. 
%%%%%%%%%%%%%%%%%%%%%%%%%%%%%%%%%%%%%%%%%%%%%%%%%%%%%%%%%%%%%%%%%%%%%
%%%%%%%%%%%%%%%%%%%%%%%%%%%% Performance Comparison %%%%%%%%%%%%%%%%%%%%%%%%%%%
%%%%%%%%%%%%%%%%%%%%%%%%%%%%%%%%%%%%%%%%%%%%%%%%%%%%%%%%%%%%%%%%%%%%%
\section{Evaluation and Analysis}
\label{sec:PerformanceComp}
In this section, we evaluate the performance of our proposed schemes. We consider that \db's covered area is modeled as a $\sqrt{\ncells}\times\sqrt{\ncells}$ grid that contains \ncells~cells each represented by one location pair (\lx,\ly) in \db. We use the efficient {\em cuckoo filter} implementation provided in~\cite{CuckooFilter} for our performance analysis with a very small false positive rate $\fp = 10^{-8}$ and a load factor $\alpha = 0.95$. In addition, since personal/portable \TVBD~devices of \su s can only transmit on available channels in the frequency bands $512-608$ MHz (TV channels $21-36$) and 614-698 MHz (TV channels $38-51$), this means that users can only access 31 white-space TV band channels in a dynamic spectrum access manner \cite{to2015electronic}. Therefore, in our evaluation we set the number of TV channels $\nch=31$.

Since in practice, at a given time, only a percentage of \db's entries contains available channels, we have ran an experiment to learn what a realistic value of this percentage might be. We denote this percentage (averaged over time and space) as \perc. We have used the Microsoft online white spaces database application~\cite{microsoft} to identify and measure \perc~by monitoring 8 different US locations (Portland, San Faransico, Houston, Miami, Seattle, Boston, New York and Salt Lake City) for few days with an interval between successive measurements of $3$ hours. Our measurements show that $\perc$ is about $6.8 \%$.

Not only does this experiment allow us to evaluate the communication overhead, but also the computational overhead, especially from the database side since both overheads are linear functions of the percentage \perc~as we show in Table~\ref{tab:Table1}.

There are several factors that influence the performance of both \LPDB~and \LPDBQS. One of these factors is the percentage \perc~which has a significant influence on the performance of our schemes as we show in Table~\ref{tab:Table1} and Figure~\ref{perc}. Also, the number of cells in the grid covered by \db~has a direct impact on the size of \db, and thus on the communication and computational overheads of \db~as highlighted in Table~\ref{tab:Table1} and Figures~\ref{fig:comm} and~\ref{fig:compDB}. In fact, as the number of cells increases, the size of \db~increases and so does the computational complexity of constructing the cuckoo filter. In addition, the false positive rate, \fp, has an impact on the cost of storing one record in the cuckoo filter and subsequently on the communication overhead as we illustrate and discuss in Figure~\ref{fig:fpsp} and Table~\ref{tab:Table1}. Finally, the fraction of positive queries, $f_p$, can impact the lookup performance as we show and discuss in Figure~\ref{fig:tput}. We discuss these factors in more details in the next section.

Next, we also compare our schemes with respect to existing approaches in terms of (i) communication and computational overhead, and (ii) location privacy. Since the schemes in~\cite{zhang2015achieving,zhang2015optimal} try to achieve a different goal, which is the mutual location privacy between \su s and \pu s, we do not include them in our overhead analysis. Note that, since the \pir~protocol used in~\cite{troja2014leveraging} has not been specified, we use the protocol proposed by Trostle et al.~\cite{trostle2010efficient} used in \PrSpec~\cite{gao2013location} in our performance comparison.

\subsection{Communication and Computation Overhead}
\label{performance}

\begin{table*}[t!]

\centering  \caption{ Communication and computation overhead of proposed and existent schemes} \label{tab:Table1}

\renewcommand{\arraystretch}{1.3}{
\resizebox{.97\textwidth}{!}{%
\begin{tabular}{||c|c||c||c|c|c||}

\hline \multicolumn{2}{||c||}{\multirow{2}{*}{\textbf{Scheme} }}   &  \multicolumn{1}{|c||}{\multirow{2}{*}{\textbf{Communication}}} & \multicolumn{3}{|c||}{\textbf{Computation}} \\ \cline{4-6}

\multicolumn{2}{||c||}{}  &  &  \textbf{ {\em DB}} & \textbf{ {\em SU}} & \textbf{ {\em QP}}\\ \hline
%\hline  \multicolumn{2}{||c||}{\textbf {\em Our Scheme w/o Precomp.}} & $ 4\gam \cdot |p| \cdot 1/2 \cdot(2+log\:n) + n\cdot \ope$  &  $1/2 \cdot(2+log\:n)\cdot 3\gam \cdot |p|\cdot Mul\gam$ & $(2\gam\cdot|p|+2\gam)\cdot Mul\gam+\ope$ & & \\ \hline
\hline  \multicolumn{2}{||c||}{$\boldsymbol{\LPDB}$} & $ \size_\qr+\perc \cdot \nch \cdot \ncells \cdot (log_2(1/\fp)+ log_2(2\beta))/\alpha$ & $\perc \cdot \nch \cdot \ncells \cdot insert$  & $\nch \cdot ( Hash + lookup)$ & - \\

\hline  \multicolumn{2}{||c||}{$\boldsymbol{\LPDB \; w/ \; leakage}$} & $ \size_\qr+\perc \cdot \nch \cdot \sqrt{\ncells} \cdot (log_2(1/\fp)+ log_2(2\beta))/\alpha$ & $\perc \cdot \nch \cdot \sqrt{\ncells} \cdot insert$  & $\nch \cdot (Hash + lookup)$ & -\\

\hline  \hline \multicolumn{2}{||c||}{\PrSpec~\cite{gao2013location}} & $ (2 \sqrt{\ncells}+ 3)\lceil \log p \rceil$ & $\mathcal{O}(\ncells)\cdot Mulp$ &  $4\sqrt{\ncells}\cdot Mulp$ & -\\

\hline \hline  \multicolumn{2}{||c||}{$\boldsymbol{\LPDBQS}$} & $ \size_\qr+\perc \cdot \nch \cdot \ncells \cdot (log_2(1/\fp)+ log_2(2\beta))/\alpha + \nch\cdot\size_\hmac$ & $\perc \cdot \nch \cdot \ncells \cdot insert$  & $\nch\cdot \hmac$ & $\nch \cdot lookup$ \\ \hline

\hline \hline  \multicolumn{2}{||c||}{Troja et al~\cite{troja2015efficient}} & $ (2 + d)\cdot log_2 \: \rsa$ & $\mathcal{O}(\ncells)\cdot Mulp$  & $4\sqrt{m\cdot v} \cdot Mulp$ & - \\ \hline

\hline \hline  \multicolumn{2}{||c||}{Troja et al~\cite{troja2014leveraging}} & $n_g\cdot b\cdot \log_2 q   + (2 \sqrt{\ncells}+ 3)\lceil \log p \rceil$ & $\mathcal{O}(\ncells)\cdot Mulp$   & $n_g\cdot b\cdot (2 Expp + Mulp) + 4\sqrt{m} \cdot Mulp$ & - \\ \hline
\end{tabular}}}

\flushleft{\scriptsize{\textbf{Variables:} $insert$ and $lookup$ denote the cost of one $Insert$ and $Lookup$ operations in the Cuckoo Filter. $\beta$ is the number of entries in a bucket of the cuckoo filter. $p$ is a large prime used in the blinding factor of \PrSpec, $q$ is a large prime used in~\cite{troja2014leveraging}, $b$ denotes the number of bits that an \su~shares with other \su s in~\cite{troja2014leveraging}, $n_g$ is the number of \su s within a same group in~\cite{troja2014leveraging}, $v$ is the size of a block in \db~\cite{troja2015efficient}, and $d$ is the umber of \db~segments in~\cite{troja2015efficient}.
$Mulp$ and $Expp$ denote a modular multiplication and a modular exponentiation operations over modulus $p$. $\size_u$ denotes the amount of data exchanged during a process $u$, where $u \in \{\qr,\hmac\}$.

}}
\end{table*}

\subsubsection{Communication Overhead}

\begin{figure}[h!]
\center
	% double column
    \includegraphics[width=0.5\textwidth]{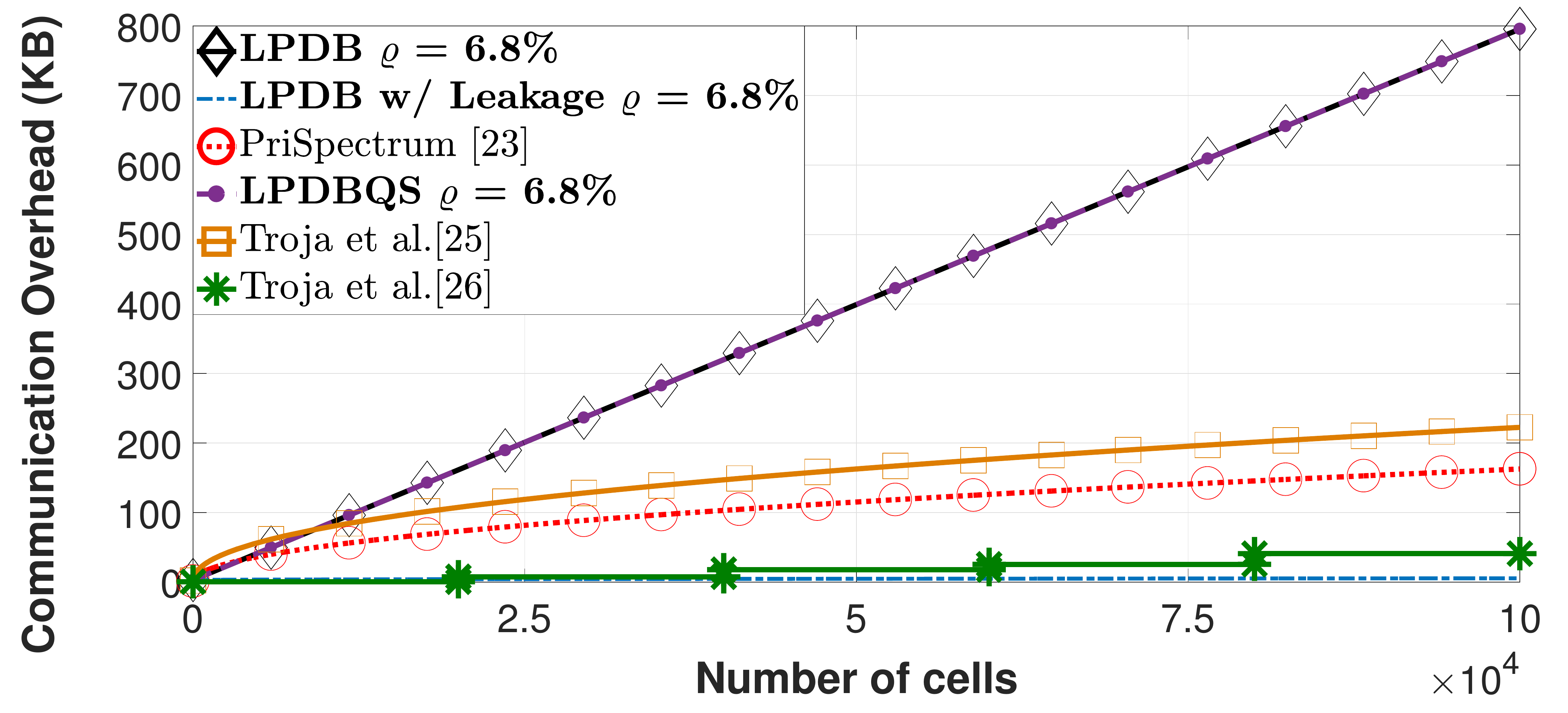}

    \caption{ Communication Overhead}
    \label{fig:comm}
\end{figure}

We provide analytical expressions of the communication overhead of these schemes in Table~\ref{tab:Table1}. For \LPDB,~we provide two expressions of the overhead with respect to two scenarios: (i) when \su s do not reveal one of their coordinates, (ii) when one of the coordinates is revealed by \su s. In both scenarios the data transmitted consist basically of \qr, sent by \su, and the response of \db~to it. The size of the response generated by \db~depends on the number of entries in \db~that satisfy \qr~and on the space needed to store each of these entities in \ckf. The number of entries for \LPDB~is given by $\perc \cdot \nch \cdot \ncells$ and reduces to $\perc \cdot \nch \cdot \sqrt{\ncells}$ when one of the coordinates is revealed by \su. $\nch \cdot \ncells$ and $\nch \cdot \sqrt{\ncells}$ provide the number of entries in \db~that satisfy the query of \su~for both scenarios. $\perc$ gives the percentage of those entries with available channels. $\LPDBQS$ incurs a slightly higher communication overhead than $\LPDB$ from a system point of view, as \su~needs to additionally send a maximum of $\nch \cdot \size_\hmac$ to \qp. However, most of this overhead is incurred between \db~and \qp~as \su s do not have to download \ckf s from \db~anymore.
For illustration purpose, we plot in Figure~\ref{fig:comm} the system communication overhead of the different schemes using the expressions established in Table~\ref{tab:Table1}.

As shown in Figure~\ref{fig:comm}, and as expected, \LPDB~is clearly more expensive than the other schemes in terms of communication overhead even when \perc, determined experimentally, is equal to $6.8\%$. However, revealing one of the coordinates brings a huge gain and makes our scheme even better than existing approaches, yet without compromising the location privacy. \LPDBQS~has almost the same communication overhead as $\LPDB$ but with the difference that most of this overhead is incurred between \db~and \qp.

\begin{figure}[h!]
 \center

 % double column
 \includegraphics[width=0.5\textwidth]{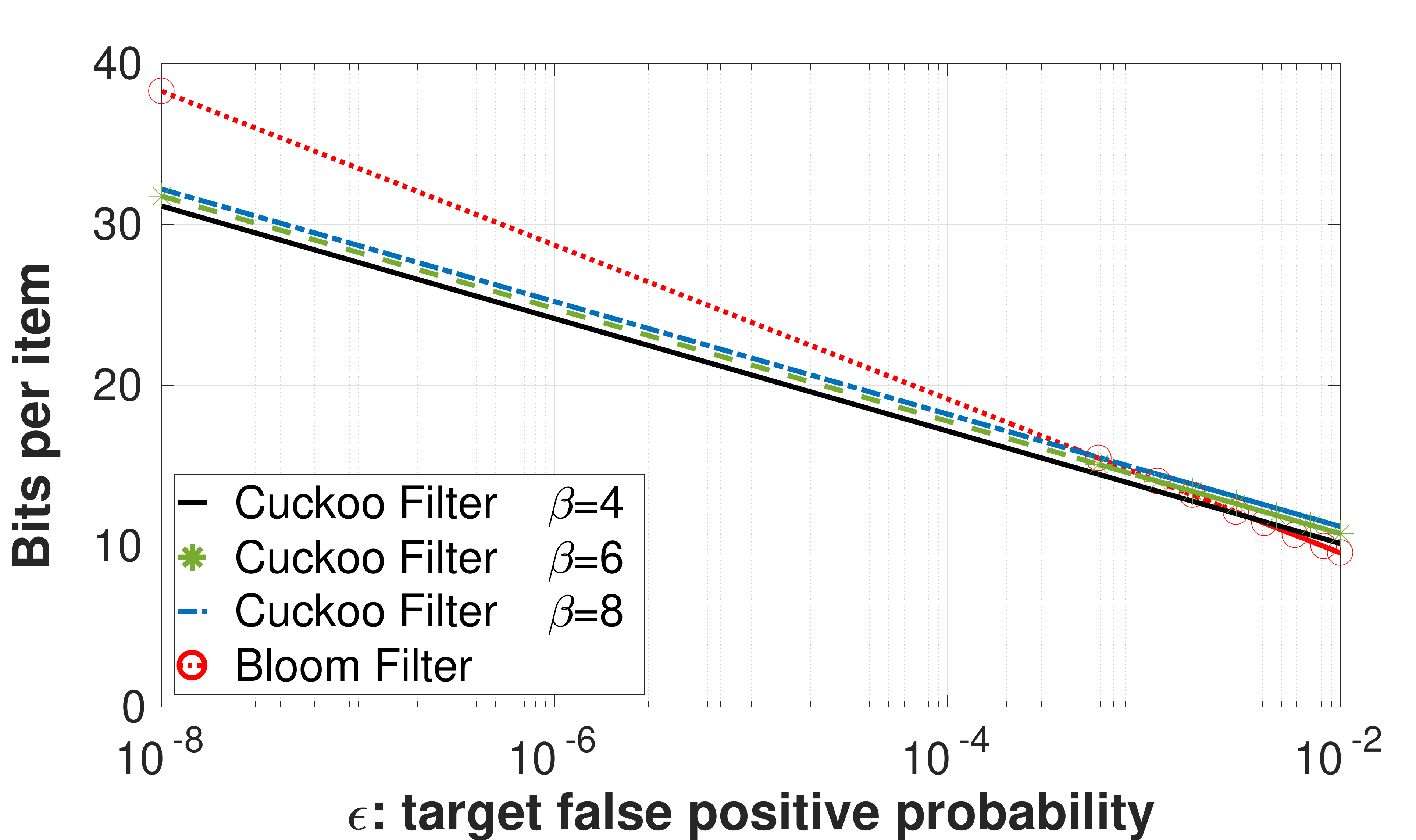} 

 \caption{False positive rate vs. space cost per element}
\label{fig:fpsp}

 \end{figure}

We study also the impact of varying the target false positive rate, $\epsilon$, on the cost of inserting one record in the \ckf~in bits as illustrated in Figure~\ref{fig:fpsp}. This has a direct impact on the size of the filter and thus the communication overhead of our schemes. We do this for multiple values of $\beta$, which is the number of slots per bucket in the cuckoo filter.  As shown in Figure~\ref{fig:fpsp}, targeting a smaller value of \fp~costs more bits to store an item in the filter and subsequently increases the communication overhead. Increasing the value of $\beta$ will require more bits per item to achieve the same target \fp~as illustrated in the Figure. However, cuckoo filter still achieves significantly better than other probabilistic data structures like space-optimized bloom filter as shown in the Figure, which again justifies our choice of the cuckoo filter technique.

\subsubsection{Computational Overhead}

 We also investigate the efficiency of our proposed schemes in terms of their computational overhead. We evaluate the computation required at each entity separately, and we provide the corresponding analytical expression of the overhead as shown in Table~\ref{tab:Table1}. Again we provide two estimated costs for both scenarios of \LPDB. The computation of \db~is given in terms of the number of insertions it has to perform into \ckf. This depends on the number of \db~entries that comply with \qr~considering only the available channels. This number is equal to $\perc \cdot \nch \cdot \ncells$ in \LPDB~and reduces to $\perc \cdot \nch \cdot \sqrt{\ncells}$ in {\em \LPDB~with leakage}. For the computational cost at the \su's side, \LPDB's overhead depends solely on the number of possible channels, \nch, and the cost of one $Hash$ and one $Lookup$ operations, as shown in Table~\ref{tab:Table1}. One of the reasons that motivated our use of the {\em cuckoo filter}, as we mentioned earlier, is that it is characterized by an extremely fast $Lookup$ operation. This allows \su s to check whether a specific combination, $\y$, exists in the filter, i.e. whether channel is available, very efficiently. \LPDB's overhead at \su's side does not depend on the size of \db~since any lookup query to \ckf~always reads a fixed number of buckets (at most two)~\cite{fan2014cuckoo}, which makes our scheme more scalable than existing approaches in terms of computation when the size of \db~increases. In \LPDBQS, \db~performs the same computation as in \LPDB. The $Lookup$ operations on \ckf~are now outsourced to \qp~instead of \su s and \qp~needs to perform a maximum of $\nch \cdot lookup$ for every querying \su, which is very fast to perform as we mentioned earlier. Every \su~needs to only construct \hmac-strings $\{y_{k_t}\}_{t=t_0}^{t_f}$ which could be done extremely quickly and could even be precomputed. Note that the \pir-based approaches have similar cost on \db's side, since in any \pir~scheme, the server is destined to have $\mathcal{O}(\ncells)$ computation~\cite{trostle2010efficient}.

\begin{figure*}[!htb]
    \centering
    \subcaptionbox{\small DB Computational Overhead.\label{fig:compDB}}
    {\includegraphics[width=0.45\textwidth]{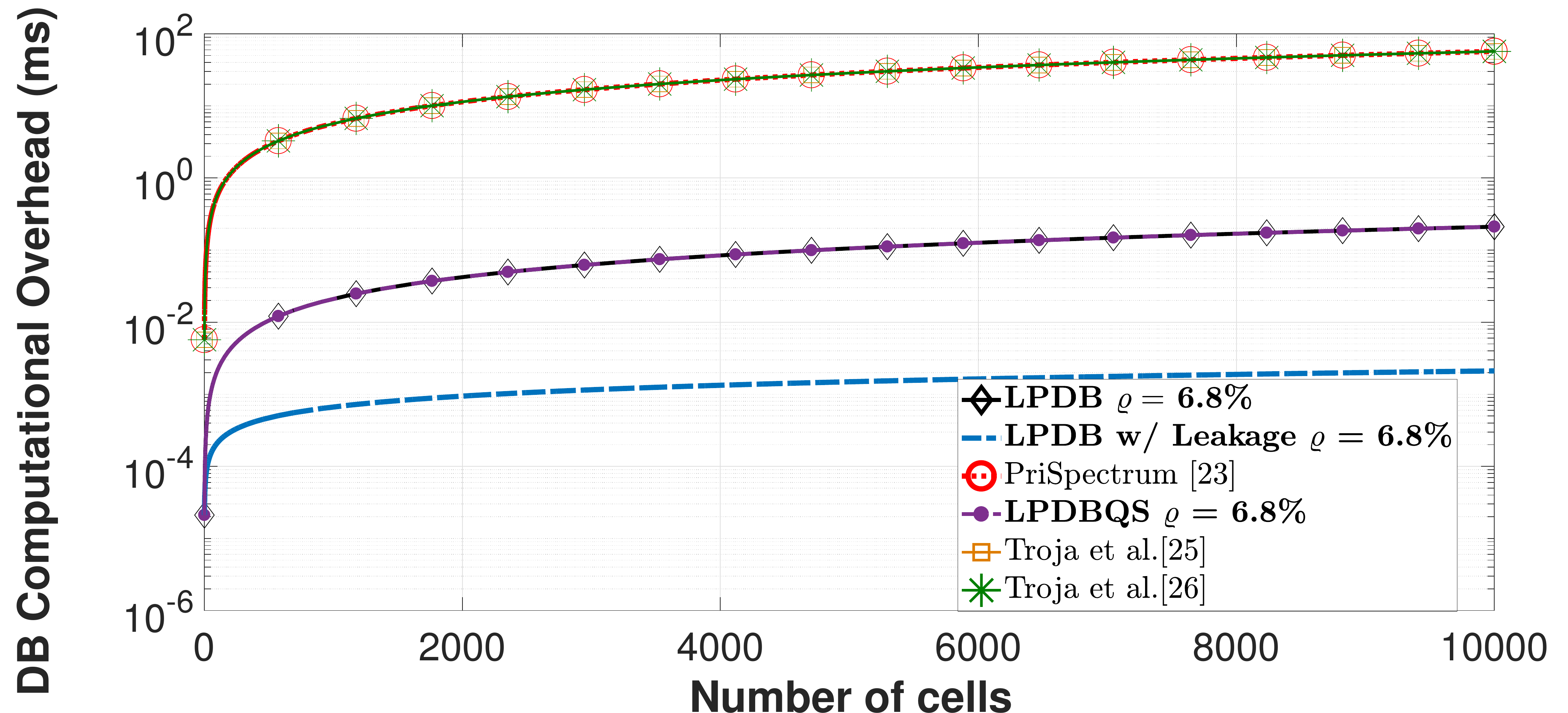}}\quad
    \subcaptionbox{ SU Computational Overhead.\label{fig:compSU}}
     {\includegraphics[width=0.45\textwidth]{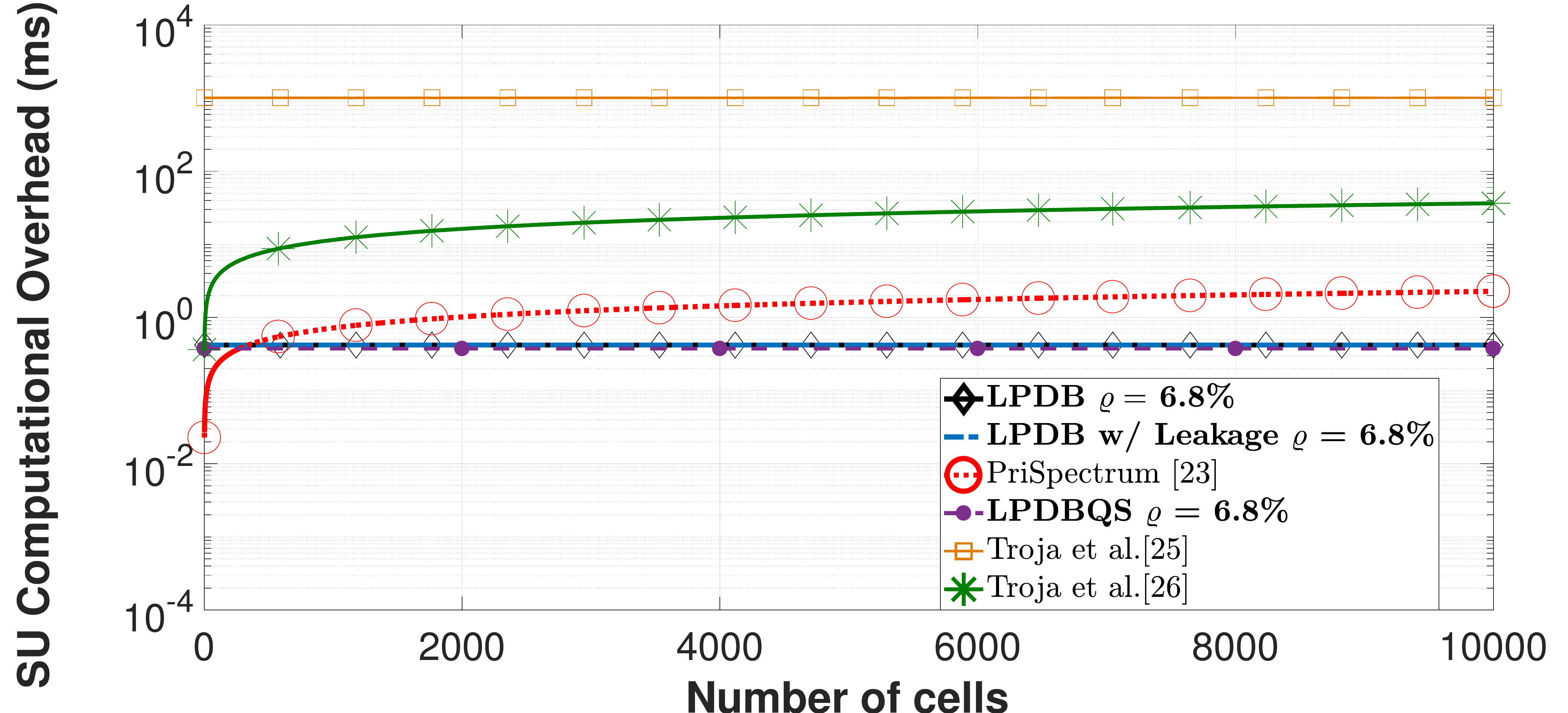}}
    \caption{ Computation Comparison}
    \label{fig:comp}
\end{figure*}

For illustration purpose, we plot in Figure~\ref{fig:comp} the computational overhead incurred by each \su~and \db, in the different schemes using the expressions established in Table~\ref{tab:Table1}.

% double column

% double column
\begin{figure*}[!htb]
\centering
    \subcaptionbox{ Communication overhead.}
    {\includegraphics[width=0.45\textwidth]{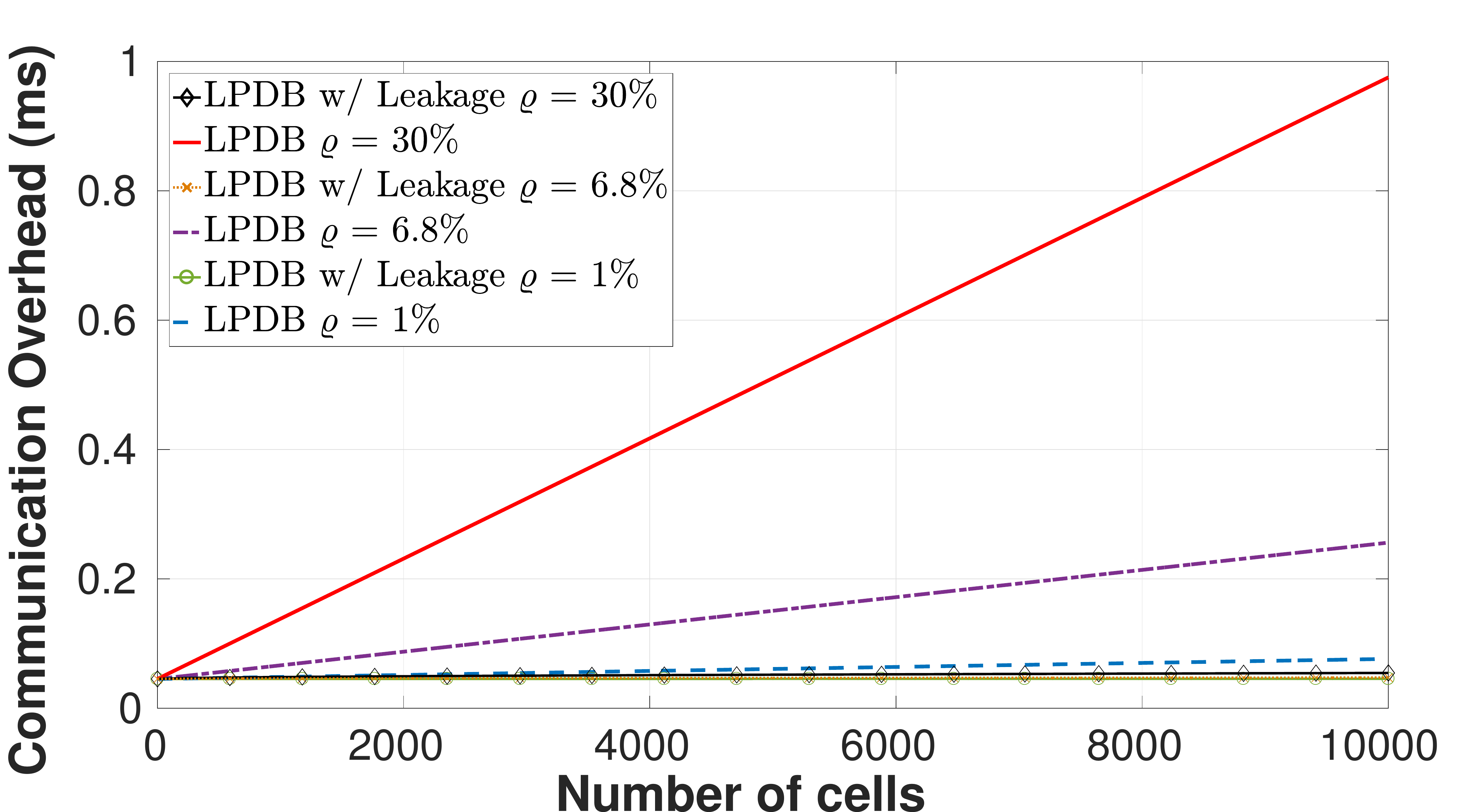}}\label{fig:commOv}\quad
    \subcaptionbox{ Computational Overhead.}
     {\includegraphics[width=0.45\textwidth]{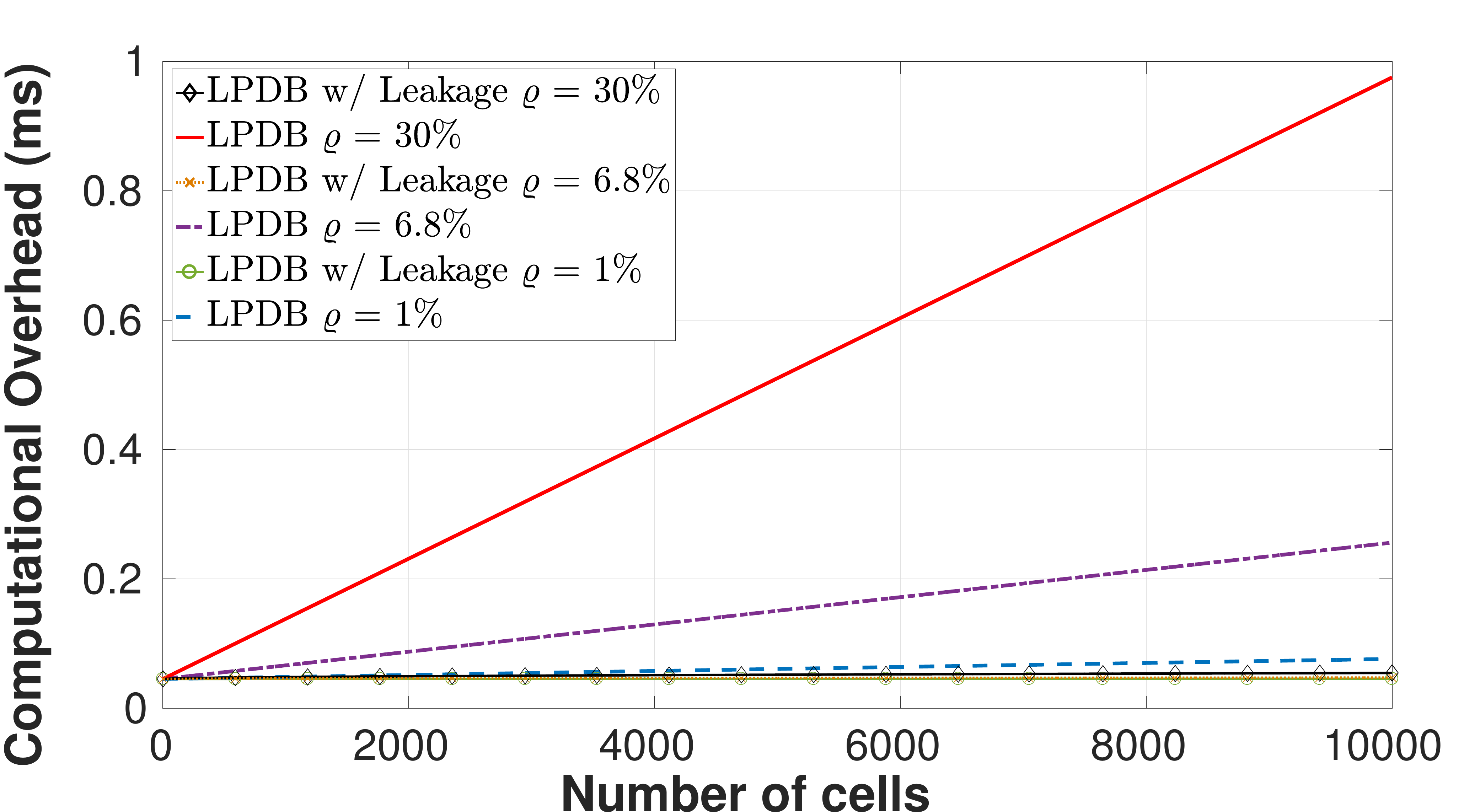}}
    \caption[]{ Impact of varying \perc.}
    \label{perc}
\end{figure*}

% doublecolumn
    \begin{figure*}[h!]
\centering   

    \begin{subfigure}[b]{0.35\textwidth}
        \includegraphics[width=\textwidth]{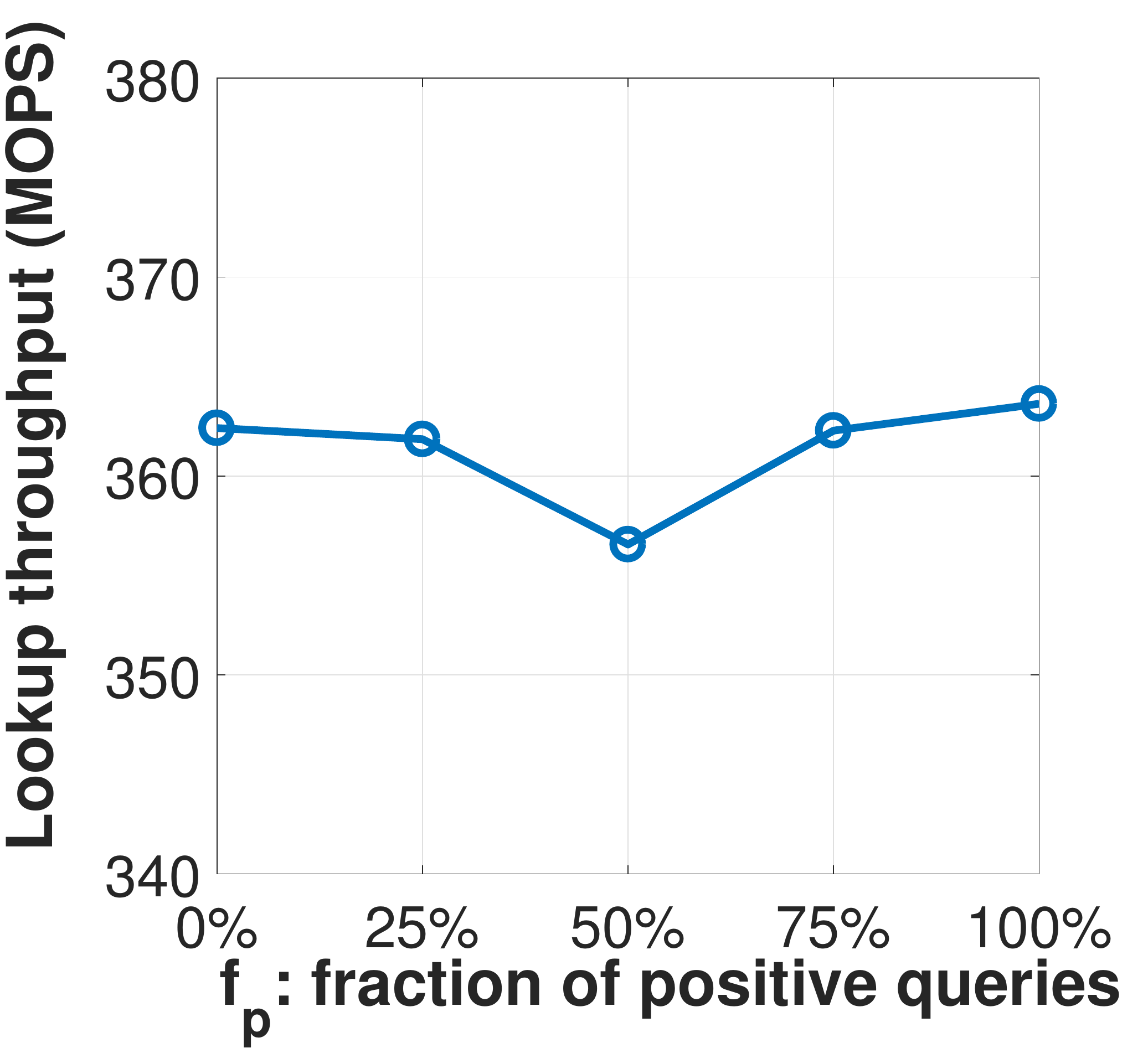}
        \caption{} \label{fig:tput}
    \end{subfigure} 
    \begin{subfigure}[b]{0.35\textwidth}
        \includegraphics[width=\textwidth]{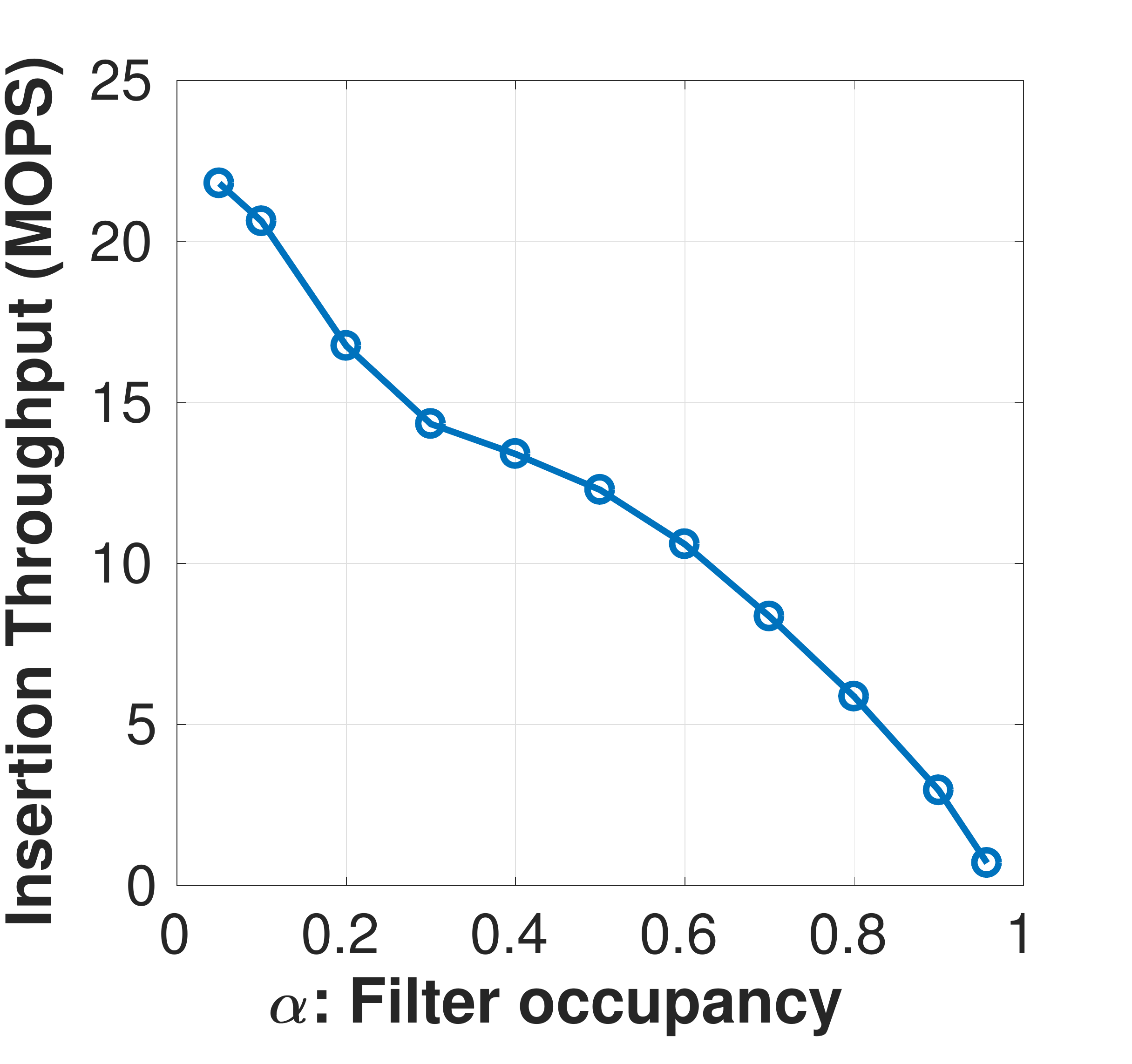}
        \caption{}\label{fig:ins}
    \end{subfigure}
    \caption{\ref{fig:tput}. Lookup performance when a filter achieves its capacity. \ref{fig:ins}. Insertion throughput for different load factors $\alpha$. }\label{fig:otherparams}

\end{figure*}

 Our schemes are much more efficient than existing approaches at both \db~and \su~sides as shown in Figures~\ref{fig:compDB} \& \ref{fig:compSU}. The gap keeps increasing considerably as the number of cells (i.e., the size of \db) increases. This is due to the fact that these approaches' cost is dominated by an increasing number of modular multiplications which are very expensive compared to the $Insert$ and $Lookup$ operations of the {\em cuckoo filter} in our schemes.

We also evaluate the impact of other parameters on the overhead perceived by both \su s and \db~as shown in Figure~\ref{fig:otherparams}. First, in Figure~\ref{fig:tput}, we illustrate the variation of the throughput of the lookup operations in million operations per second (MOPS) in a cuckoo filter of size $112 MB$ as a function of the fraction of positive queries $f_p$, i.e. queries for items that actually exist in the filter. This clearly shows the efficiency of the lookup operations that \su~or \qp~has to perform to check availability information within \ckf. \ckf~always fetches two buckets and thus achieves about the same performance when the queries are $100\%$ positive or $100\%$ negative and drops when $f_p = 50\%$ for which the CPU's branch prediction is least accurate~\cite{fan2014cuckoo}.

 We also assess the insertion throughput that \db~experiences to construct the \ckf~as a function of the load factor $\alpha$ as shown in Figure~\ref{fig:ins}. As opposed to the lookup throughput shown in Figure~\ref{fig:tput}, \ckf~has a decreasing insert throughput when it is more filled (though their overall construction speed is still high). This is mainly due to the fact that \ckf~may have to move a sequence of existing fingerprints recursively before successfully inserting a new item, and this process becomes more expensive when the load factor grows higher~\cite{fan2014cuckoo}.

\subsubsection{Impact of varying the percentage \perc~of entries with available channels}

We also study the impact of \perc~on the overhead incurred by our schemes. For this, we plot in Figure~\ref{perc} the communication and the system computational overheads for different values of \perc. We plot only \LPDB~and {\em \LPDB~with leakage} as $\LPDBQS$ has almost the same overhead as \LPDB. As shown in Figure~\ref{perc}, both overheads behave similarly in the way that decreasing \perc~when one of the coordinates is revealed doesn't impact much our scheme. {\em \LPDB~w/ Leakage} has the smallest overhead compared to the case where no leakage is allowed. On the other hand, decreasing this parameter drastically reduces the overhead of \LPDB~and even makes it comparable to {\em \LPDB~w/ Leakage} in terms of communication and computation. This means that in the case where only $1\%$ or less of \db~entries have available channels, there is no need to reveal one of the coordinates to reduce the overhead.

\subsection{Location privacy}

We compare our schemes to existing approaches in terms of location privacy level by presenting the security problems on which they rely as illustrated in Table~\ref{tab:locpr}. We also precise the localization probability of \su s under these schemes. The best probability that could be achieved is $1/m$, i.e. \su s are within \db~coverage area. If one of the schemes is broken then this probability increases considerably.
\begin{table}[h!]

\centering  \caption{ Location privacy} \label{tab:locpr}

\renewcommand{\arraystretch}{1.3}{
\resizebox{.49\textwidth}{!}{%
\begin{tabular}{||c|c||c||c||}

\hline \multicolumn{2}{||c||}{\textbf{Scheme} }   &  \multicolumn{1}{|c||}{\textbf{Security level}} &  \multicolumn{1}{|c||}{\textbf{Localization probability}}  \\  \hline

\hline  \multicolumn{2}{||c||}{$\boldsymbol{\LPDB}$} & Unconditionally secure & $1/m$ \\

\hline  \multicolumn{2}{||c||}{$\boldsymbol{\LPDB \; w/ \; leakage}$} &  Uncond. within 1 coordinate &$\sqrt{1/m}$ \\

\hline \hline \multicolumn{2}{||c||}{\PrSpec~\cite{gao2013location}} & Computational \pir & $1/m$ \\

\hline \multicolumn{2}{||c||}{Troja et al~\cite{troja2015efficient}} & Computational \pir & $1/m$ \\

\hline \multicolumn{2}{||c||}{Troja et al~\cite{troja2014leveraging}} & Computational \pir  & $1/m$\\

\hline \hline  \multicolumn{2}{||c||}{$\boldsymbol{\LPDBQS}$} & $\kappa-\hmac$ & $ 1/m$  \\

\hline \hline  \multicolumn{2}{||c||}{Zhang et al.~\cite{zhang2015optimal}} & {\em $k$-anonymity} & $1/k$  \\

\hline  \multicolumn{2}{||c||}{Zhang et al.~\cite{zhang2015achieving}} & Geo-Indistinguishability & $1/r$  \\ \hline
\end{tabular}}}
\flushleft{\center \scriptsize{\textbf{Variables:} $r$ is the radius of the $\epsilon$-geo-indistinguishability mechanism in~\cite{zhang2015achieving}.

}}
\end{table}

\LPDB~offers unconditional security, as \su s do not share any information that could reveal their location. \LPDB~could be seen as a variant of \pir~in which the server sends a whole copy of the database to the user and this is the only way to achieve information theoretic privacy (i.e. cannot be broken even with computationally unbounded adversary) in a single-server setting. Even if one of the coordinates is intentionally revealed by a \su, its location is still indistinguishable from $\sqrt{m}-1$ remaining possible locations.

The approaches in~\cite{gao2013location,troja2015efficient,troja2014leveraging} rely on computational \pir~protocols to preserve \su s' location privacy. The security of Computational \pir~protocols' is established against a computationally bounded adversary based on well-known cryptographic problems that are hard to solve (e.g. discrete logarithm or factorization~\cite{menezes1996handbook}). This means that these approaches offer lower security level than \LPDB.

%For instance, \PrSpec~\cite{gao2013location} uses the computational \pir~of Trostle et al.~\cite{trostle2010efficient}. This \pir~relies on the hard to solve problem of finding the group order used in Merkle-Hellman cryptosystem which is based on the knapsack problem. Troja et al.~\cite{troja2015efficient} use the computational \pir~of Gentry et al.~\cite{gentry2005single} whose security is based on the {\em $\phi$-hiding} assumption~\cite{cachin1999computationally}.

The approach proposed by Zhang et al.~\cite{zhang2015optimal} relies on the concept of {\em k-anonymity}, which offers very low privacy level as the probability of identifying the location of a querying \su~is equal to $1/k$. Also, an approach cannot be proved to satisfy {\em k-anonymity} unless assumptions are made about the attacker's auxiliary
information. For instance, dummy locations are only
useful if they look equally likely to be the real location from
the adversary's point of view. Any auxiliary information that allows the attacker to rule out any of those locations would immediately
violate the definition.

As we have shown in Section~\ref{sec:secAnalysis}, \LPDBQS~is as secure as its underlying \hmac~which is breakable only with probability of $1/2^\kappa$, where $\kappa$~is the security level. For the same security level, \hmac~incurs much less communication overhead than that of the computational \pir~protocols in~\cite{gao2013location,troja2015efficient,troja2014leveraging}.

Zhang et al.~\cite{zhang2015achieving} propose an approach whose privacy depends on the $\epsilon$-geo-indistinguishability~\cite{andres2013geo} mechanism, which is derived from the {\em differential privacy} concept. In this mechanism, a \su~sends a randomly chosen point $z$ close to its location, but that still allows it to get a useful service. An informal, definition of this mechanism as given in~\cite{andres2013geo} is as follows: A mechanism satisfies $\epsilon$-geo-indistinguishability if and only if for any radius $r > 0$, the user enjoys $\ell$-privacy within a radius $r$, where $\ell = \epsilon r$ and $\epsilon$ is the privacy level per unit of distance. A user is said to enjoy $\ell$-privacy within $r$ if, by observing $z$, the adversary's ability to find the user's location among all points within $r$, does not increase by more than a factor depending on $\ell$ compared to the case when $z$ is unknown~\cite{andres2013geo}. The smaller $\ell$ the stronger the privacy the user enjoys. \su~can specify its privacy level requirement by providing the radius $r$ it is concerned about, and the privacy level that it wishes for this specific radius. Relying on this mechanism in the context of \crn, is problematic because, first, it introduces some noise to \su's location which may cause erroneous spectrum availability information and, subsequently, interference with primary transmissions. Second, to avoid facing the previous issue, \su~may need to pick the radius that can still give it accurate information which means necessarily that $r << \sqrt{\ncells}$. Hence, even though the adversary will be unable to pinpoint the exact location of the \su, it will be able though to learn that it is within the radius $r$ from the the shared location $z$.

In summary, as can be seen in Table~\ref{tab:locpr} and as explained above, \LPDB~offers the highest location privacy level as it achieves information-theoretic security. \LPDBQS~can offer similar security guarantees as computational \pir-based approaches but with significantly better computational and communication overhead thanks to the use of \hmac.
 
%%%%%%%%%%%%%%%%%%%%%%%%%%%%%%%%%%%%%%%%%%%%%%%%%%%%%%%%%%%%%%%%%%%%%
%%%%%%%%%%%%%%%%%%%%%%%%%%%% Conclusion %%%%%%%%%%%%%%%%%%%%%%%%%%%
%%%%%%%%%%%%%%%%%%%%%%%%%%%%%%%%%%%%%%%%%%%%%%%%%%%%%%%%%%%%%%%%%%%%%

\section{Conclusion}
\label{sec:Conclusion}
In this paper, we have proposed two location privacy preserving schemes, called \LPDB~and \LPDBQS, that aim to preserve the location privacy of \su s in database-driven $\crn$s. They both use {\em set membership data structures} to transmit a compact representation of the geo-location database to either \su~or \qp, so that \su~can query it to check whether a specific channel is available in its vicinity. These schemes require different architectural and performance tradeoffs.
\section*{Acknowledgment}
This work was supported in part by the US National Science Foundation under NSF award CNS-1162296.

%%%%%%%%%%%%%%%%%%%%%%%%%%%%%%%%%%%%%%%%%%%%%%%%%%%%%%%%%%%%%%%%%%%%%
%%%%%%%%%%%%%%%%%%%%%%%%%%%%  References %%%%%%%%%%%%%%%%%%%%%%%%%%%%
%%%%%%%%%%%%%%%%%%%%%%%%%%%%%%%%%%%%%%%%%%%%%%%%%%%%%%%%%%%%%%%%%%%%%
\small{
\bibliographystyle{IEEEtran}
\bibliography{IEEEabrv,./references}
\vspace{-35pt}
}

\begin{IEEEbiography}[{\includegraphics[width=1in,height=1.25in,clip,keepaspectratio]{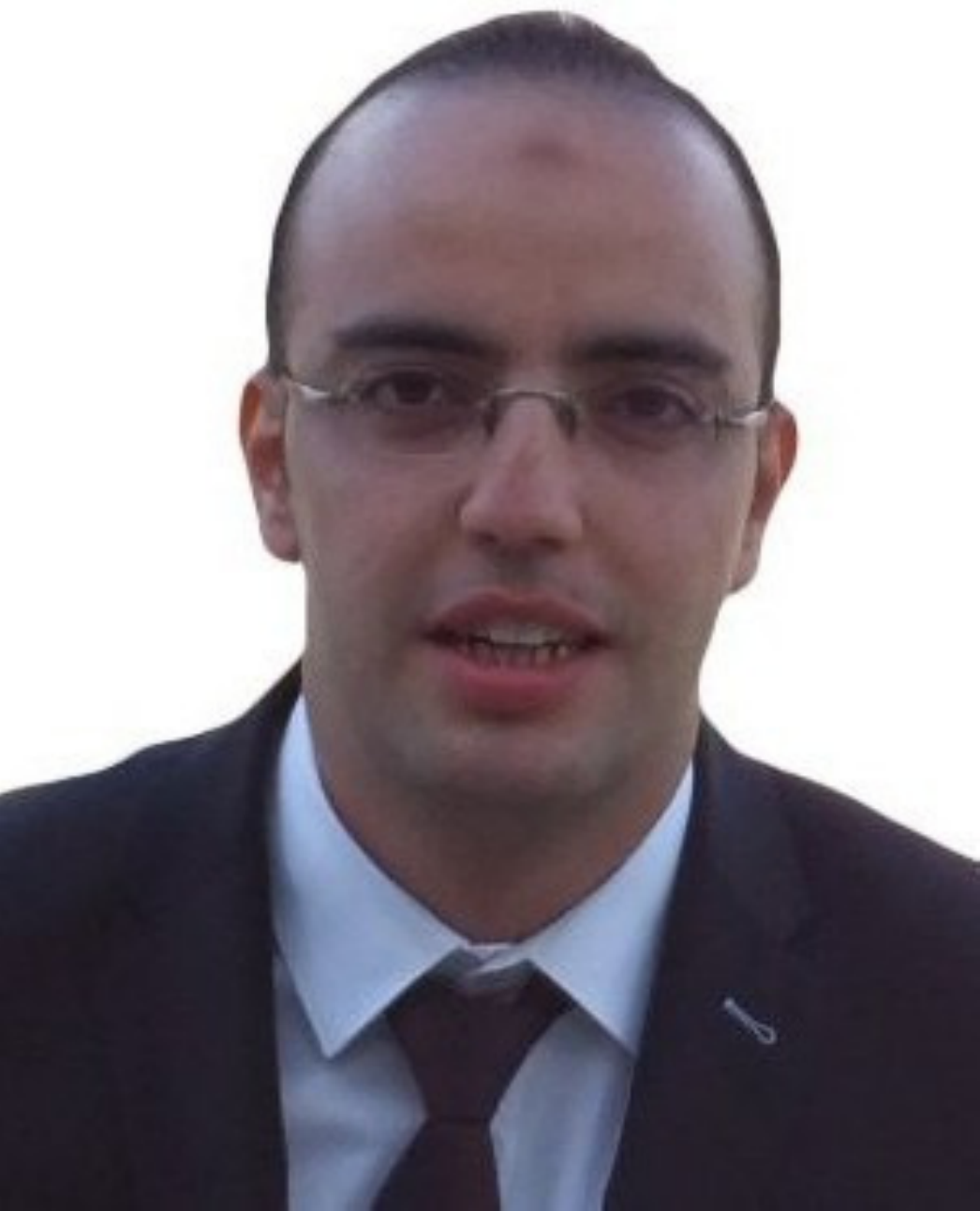}}]{Mohamed Grissa}(S'14) received the Diploma of Engineering (with highest distinction)
in telecommunication engineering from Ecole
Superieure des Communications de Tunis, Tunis,
Tunisia, in 2011, and the
M.S. degree in electrical and computer engineering
(ECE) from Oregon State University, Corvallis, OR, USA, in 2015. He is currently working
toward the Ph.D. degree at the School of Electrical
Engineering and Computer Science (EECS), Oregon
State University, Corvallis, OR, USA.

Before pursuing the Ph.D. degree, he worked as a Value Added Services Engineer at Orange France Telecom Group from 2012 to 2013. His research interests include privacy and security in wireless networks, cognitive radio networks, IoT and eHealth systems.
\vspace{-35pt}
\end{IEEEbiography}

\begin{IEEEbiography}[{\includegraphics[width=1in,height=1.25in,clip,keepaspectratio]{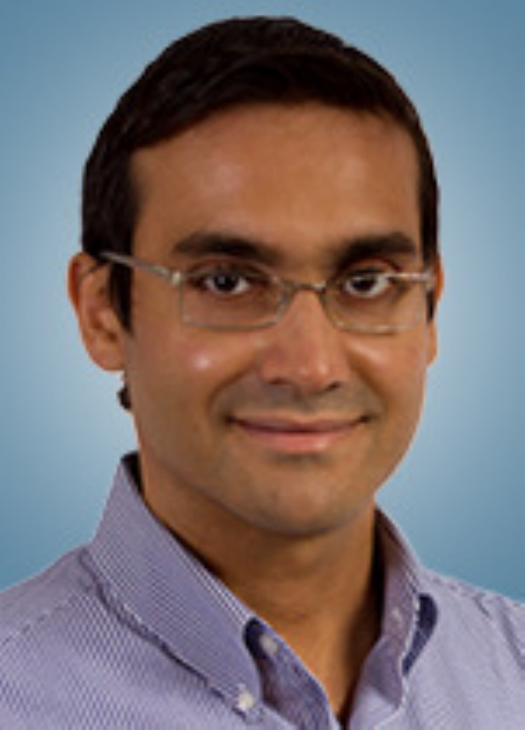}}]{Attila A. Yavuz} (S'05\textendash M'10) received a BS degree in Computer Engineering from Yildiz Technical University (2004) and a MS degree in Computer Science from Bogazici University (2006), both in Istanbul, Turkey. He received his PhD degree in Computer Science from North Carolina State University in August 2011. Between December 2011 and July 2014, he was a member of the security and privacy research group at the Robert Bosch Research and Technology Center North America. Since August 2014, he has been an Assistant Professor in the School of Electrical
Engineering and Computer Science, Oregon State University, Corvallis, USA. He is also an adjunct faculty at the University of Pittsburgh's School of Information Sciences since January 2013.

Attila A. Yavuz is interested in design, analysis and application of cryptographic tools and protocols to enhance the security of computer networks and systems. His current research focuses on the following topics: Privacy enhancing technologies (e.g., dynamic symmetric and public key based searchable encryption), security in cloud computing, authentication and integrity mechanisms for resource-constrained devices and large-distributed systems, efficient cryptographic protocols for wireless sensor networks.
\vspace{-35pt}
\end{IEEEbiography}

\begin{IEEEbiography}[{\includegraphics[width=1in,height=1.25in,clip,keepaspectratio]{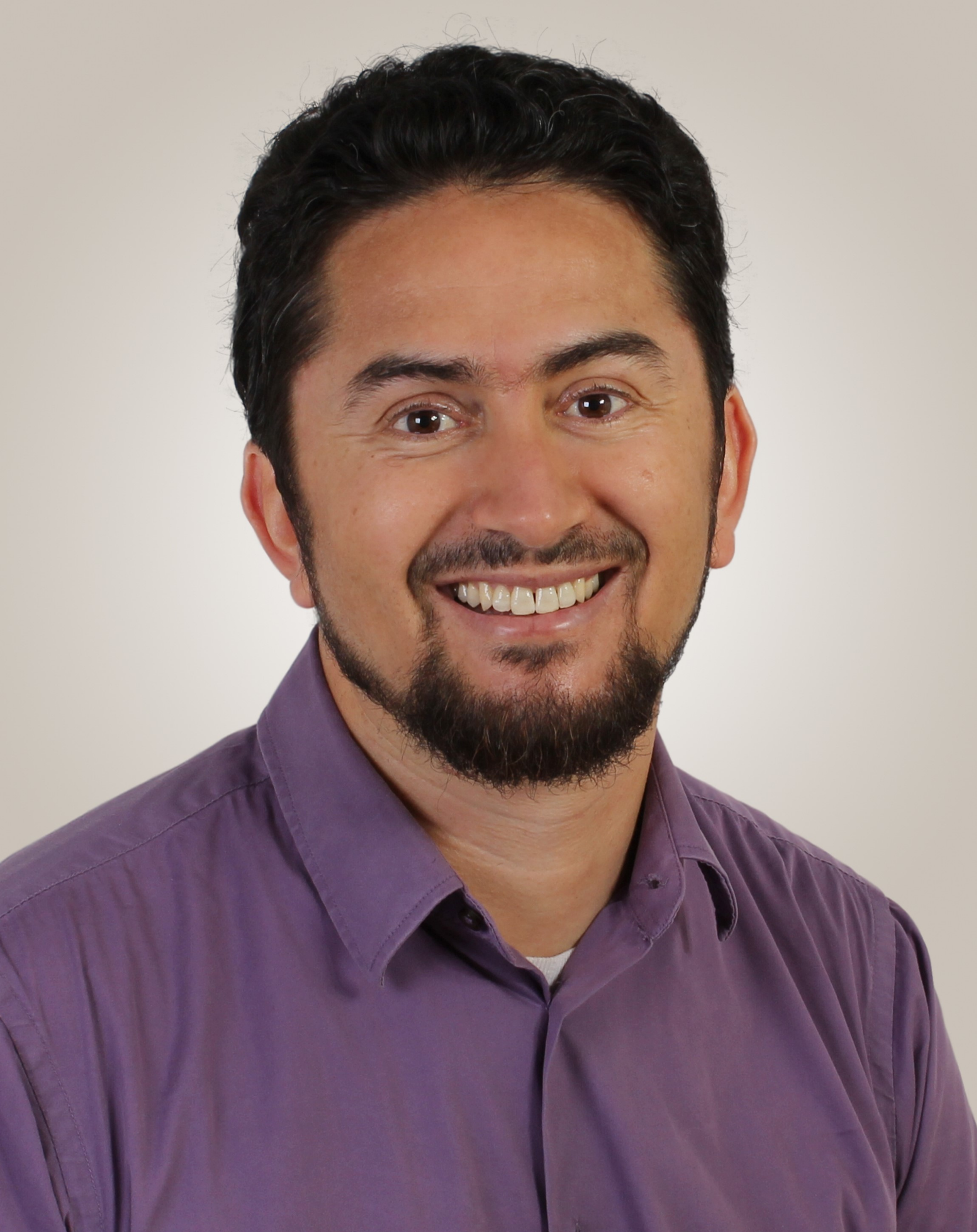}}]{Bechir Hamdaoui} (S'02\textendash M'05\textendash SM'12) is presently an Associate Professor in the School of EECS at Oregon State University. He received the Diploma of Graduate Engineer (1997) from the National School of Engineers at Tunis, Tunisia. He also received M.S. degrees in both ECE (2002) and CS (2004), and the Ph.D. degree in ECE (2005) all from the University of Wisconsin-Madison. His research interest spans various areas in the fields of computer networking, wireless communications, and mobile computing, with a current focus on distributed optimization, parallel computing, cognitive networks, cloud computing, and Internet of Things. He has won several awards, including the 2016 EECS Outstanding Research Award and the 2009 NSF CAREER Award. He is presently an Associate Editor for IEEE Transactions on Wireless Communications (2013-present). He also served as an Associate Editor for IEEE Transactions on Vehicular Technology (2009-2014), Wireless Communications and Mobile Computing Journal (2009-2016), and for Journal of Computer Systems, Networks, and Communications (2007-2009). He is currently serving as the chair for the 2017 IEEE INFOCOM Demo/Posters program. He has also served as the chair for the 2011 ACM MOBICOM's SRC program, and as the program chair/co-chair of several IEEE symposia and workshops, including GC 16, ICC 2014, IWCMC 2009-2017, CTS 2012, and PERCOM 2009. He also served on technical program committees of many IEEE/ACM conferences, including INFOCOM, ICC, and GLOBECOM. He has been selected as a Distinguished Lecturer for the IEEE Communication Society for 2016 and 2017. He is a Senior Member of IEEE, IEEE Computer Society, IEEE Communications Society, and IEEE Vehicular Technology Society.
\end{IEEEbiography}

\end{document}